\documentclass{aa}  

\usepackage{lscape}
\usepackage{graphicx}
\usepackage{txfonts}
\begin{document}

   \title{Searching for chemical signatures of brown dwarf formation
     \thanks{
      Based on observations made with the Mercator Telescope;
      on observations made with the Nordic Optical Telescope;
      on data products from the  SOPHIE archive; on data products
      from the ELODIE archive; 
      and on data products from observations made with ESO Telescopes
      at the La Silla Paranal Observatory under programme ID 
      072.C-0488(E),
 076.C-0155(A),
 076.C-0429(A),
 078.C-0133(A),
 079.C-0329(A),
 082.C-0333(A),
 083.C-0174(A),
 083.C-0413(A),
 085.C-0019(A),
 085.C-0393(A), 087.A-9029(A),
 087.C-0831(A),  
 090.C-0421(A),
 093.C-0409(A),
 094.D-0596(A), 095.A-9029(C), 178.D-0361(B), 
 183.C-0972(A),
 184.C-0639(A),
 and 188.C-0779(A).
     }\fnmsep
   \thanks{
     Tables~\ref{parameters_table_full}, and ~\ref{abundance_table_full},
     are only available in electronic
     format.
     }
     }

    \author{J. Maldonado
          \inst{1}
          \and  E. Villaver
          \inst{2}
          }

             \institute{INAF - Osservatorio Astronomico di Palermo,
              Piazza del Parlamento 1, I-90134 Palermo, Italy
             \and  
              Universidad Aut\'onoma de Madrid, Dpto. F\'isica Te\'orica, M\'odulo 15,
              Facultad de Ciencias, Campus de Cantoblanco, 28049 Madrid, Spain
             }

   \offprints{J. Maldonado \\ \email{jmaldonado@astropa.inaf.it}}
   \date{Received ...; Accepted ....}

 
  \abstract
   { 
    Recent studies 
    have shown that close-in brown dwarfs in the mass range 
    35-55 M$_{\rm Jup}$ are almost depleted as companions to stars,
    suggesting that 
    objects with masses 
    above and below this 
    gap might have different formation mechanisms. 
   }
   {
    We aim to test whether stars harbouring ``massive'' brown dwarfs and
    stars with ``low-mass'' brown dwarfs show any chemical peculiarity that
    could be related to different formation processes. 
   }
   {
    Our methodology is based on the analysis of high-resolution \'echelle spectra
    (R $\sim$ 57000) from 2-3 m class telescopes. 
    We determine the fundamental stellar parameters, as well as individual 
    abundances of C, O, Na, Mg, Al, Si, S, Ca, Sc, Ti, V, Cr, Mn, Co, Ni, and Zn
    for a large sample of stars known to have a substellar companion in the
    brown dwarf regime.  
    The sample is divided into stars hosting massive and low-mass brown dwarfs.
    Following previous works 
    a threshold of 42.5 M$_{\rm Jup}$ was considered.  
    The metallicity and abundance trends of both subsamples are
    compared and set in the context of current models of planetary and brown dwarf
    formation. 
    }
   {
    Our results  confirm that stars with brown dwarf companions do not follow the well-established
    gas-giant planet metallicity correlation seen in main-sequence planet hosts.
    Stars harbouring  ``massive'' brown dwarfs
    show similar metallicity and abundance distribution as stars without known planets
    or with low-mass planets.
     We find a tendency of stars harbouring ``less-massive'' brown dwarfs of having slightly
     larger metallicity, [X$_{\rm Fe}$/Fe] values, and abundances of
     Sc~{\sc ii}, Mn~{\sc i},  and Ni~{\sc i}
    in comparison with the stars having the massive brown dwarfs.
    The data suggest, as previously reported, that massive and low-mass brown
    dwarfs might present differences in period and eccentricity. 
}
  {
   We find evidence of a non-metallicity dependent mechanism for the formation of massive brown dwarfs. 
   Our results agree with a scenario in which massive brown dwarfs
   are formed as stars.
   At high-metallicities, the core-accretion mechanism might become efficient in the formation
   of low-mass brown dwarfs while at lower metallicities
   low-mass brown dwarfs  could form by
   gravitational instability
   in turbulent protostellar discs. 
}
  \keywords{techniques: spectroscopic - stars: abundances -stars: late-type -stars: planetary systems}

  \maketitle

%
\section{Introduction}

 Understanding 
 whether 
 brown dwarfs and giant planets share similar formation mechanisms 
 is the subject of intensive studies 
 \citep[e.g.][]{2007prpl.conf..443L,2007prpl.conf..459W,2011ASPC..450..113B,2012ARA&A..50...65L,2014prpl.conf..619C}.

 The standard definition of  a brown dwarf includes objects in a wide
 range of masses, from 13 to 80 Jupiter masses, with sufficient mass 
 to ignite deuterium but below the hydrogen-burning minimum
 mass \citep[][]{1997ApJ...491..856B,2001RvMP...73..719B,2000ARA&A..38..337C,2011ApJ...727...57S}. 
 It is now well-established that
 there is a paucity of close brown dwarfs companions in comparison with gas-giant 
 planets or binaries around main-sequence stars
 \citep{1988ApJ...331..902C,1993ApJ...413..349M,2000PASP..112..137M,2006ApJ...640.1051G,2011A&A...525A..95S},
 usually known as the brown dwarf desert.

 There have been several studies with the goal of understanding whether the properties
 of the brown dwarf population could be related to the formation mechanism of these objects. 
 In a recent work, \cite{2014MNRAS.439.2781M} compare the 
 orbital properties (period and eccentricities) of a sample of 
 brown dwarf companions around 65  stars. 
 They found that while brown dwarfs with minimum masses greater
 than $\sim$ 42.5 M$_{\rm Jup}$ follow a similar period-eccentricity
 distribution to that of stellar binaries, brown dwarfs with masses below
 42.5 M$_{\rm Jup}$ have an eccentricity distribution consistent with that
 of massive planets. 
 This suggests that the standard definition of brown dwarf
 might mix two kind of objects with different formation mechanisms.
 The formation of high-mass brown dwarfs might be a 
 scaled-down version of star formation 
 through fragmentation of molecular clumps.
 On the other hand, less-massive brown dwarfs might 
 form like giant-planets.

 Current models of giant-planet formation can be divided
 into two broad categories:
 i) core-accretion models \citep[e.g.][]{1996Icar..124...62P,2003ApJ...598L..55R,
 2004A&A...417L..25A,2012A&A...541A..97M}
 which are able
 to explain the observed gas-giant planet metallicity
 correlation \citep[e.g.][]{1997MNRAS.285..403G,2004A&A...415.1153S,2005ApJ...622.1102F}
 as well as the lack of a metallicity correlation in low-mass planet
 hosts  \citep[e.g.][]{2011arXiv1109.2497M,2012Natur.481..167C,2013Natur.503..381H};
 and ii) disc instability models which do no depend on the metallicity of the
 primordial disc \citep{1997Sci...276.1836B,2002ApJ...567L.149B,2006ApJ...643..501B}.
 If 
 brown dwarfs form like giant-planets and
 those are mainly formed by core-accretion,
 stars hosting 
 brown dwarfs 
 should 
 show the metal-enrichment
 seen in gas-giant planet hosts.

 Several attempts to understand the metallicity distribution
 of stars with brown dwarfs companions have been performed.
 \cite{2011A&A...525A..95S} notice that while some stars with brown
 dwarf companions are metal rich, others show sub-solar metallicities. 
 \cite{2014MNRAS.439.2781M} do not find significant metallicity
 differences between brown dwarf host stars with (minimum)
 masses below and above 42.5 M$_{\rm Jup}$.
 \cite{2014A&A...566A..83M} analyse in a homogeneous way 
 the abundances of 15 stars hosting brown dwarfs (7 ``candidates'' and 8 
 ``discarded'' based on their {\sc Hipparcos} astrometry)
 showing that they differ from those of stars hosting 
 gas-giant planets. Also, they suggest higher abundances
 for the stars hosting brown dwarfs with masses below
 42.5 M$_{\rm Jup}$.

 Given that previous works are based on small or inhomogeneous samples
 a detailed chemical analysis of a homogeneous and large
 sample of stars hosting brown dwarfs is needed before 
 formation mechanisms of 
 brown dwarfs are invoked.
 This is the goal of this paper, in which we present a homogeneous analysis of a large sample of
 brown dwarf hosts that is based on high resolution and high signal-to-noise ratio (S/N) \'echelle spectra.

 This paper is organised as follows. Section~\ref{observa}
 describes the stellar samples analysed in this work, the spectroscopic observations,
 and how stellar parameters and abundances
 are obtained. The comparison of the properties and abundances
 of stars with brown dwarf companions with masses larger and below 42.5 M$_{\rm Jup}$
 is presented in  Sect.~\ref{results_sect},
  where we also include a search for correlations between the stellar and brown dwarf properties.	 
 The results are discussed at length in Sect.~\ref{discusion}.
 Our conclusions follow in Sect.~\ref{conclusions}.

\section{Observations}
\label{observa}
\subsection{The stellar sample}

  A sample of stars with known brown dwarfs companions (SWBDs)
 with (projected) masses between 10 and 70 M$_{\rm Jup}$
 was built using as reference the 65 stars listed in the recent
 compilation by \cite{2014MNRAS.439.2781M} plus the 61 stars with brown dwarfs
 candidates listed
 by \cite{2016A&A...588A.144W}.
 Although different authors might have different criteria to classify an object as a brown dwarf, 
 we note that $\sim$ 64\% of the stars listed in \cite{2016A&A...588A.144W} were already
 given in the compilation by \cite{2014MNRAS.439.2781M}. Fifteen brown dwarf
 companions listed in \cite{2016A&A...588A.144W} were published after \cite{2014MNRAS.439.2781M}. 
 Only seven stars listed in \cite{2016A&A...588A.144W} and known
 before \cite{2014MNRAS.439.2781M} were not included in this
 compilation, five of them have projected masses
 $\sim$ 10-11 M$_{\rm Jup}$, while the other two are in the range 62-65 M$_{\rm Jup}$.
  Although we do not know the reason why these seven stars were not included
  in \cite{2014MNRAS.439.2781M} we 
  we have decided to keep them in our analysis. Thus, from the above compilations,
  we selected all stars with spectral type
 between F6 and K2 (independently of its luminosity class)
 with high-resolution spectra available in public archives
 or already observed by our team in our previous programmes
 (see below).
 Several stars having a very  low
 signal-to-noise (S/N) ratio or showing indications
 of high rotation were also discarded.

 Our final sample consists of 53 stars with brown dwarfs, 
 including 10 F stars, 31 G stars, and
 12 K stars. Regarding their evolutionary stage, 8 stars are red giants,
 19 are classified as subgiants, while 26 stars are on the
 main-sequence. The stars are listed in Table~\ref{parameters_table_full}. 
  
\subsection{Spectroscopic data}

 High-resolution spectra of the stars were mainly collected from public
 archives: data for six stars  were taken from the ELODIE \citep{1996A&AS..119..373B}
 archive\footnote{http://atlas.obs-hp.fr/elodie/},
  twenty stars from the SOPHIE \citep{2006tafp.conf..319B} archive\footnote{http://atlas.obs-hp.fr/sophie/},
 HARPS \citep{2003Msngr.114...20M} spectra from the ESO archive\footnote{http://archive.eso.org/wdb/wdb/adp/phase3\_spectral/form?phase3\_\\collection=HARPS}  
 was used for sixteen stars,
 while for three stars FEROS spectra were used \citep{1999Msngr..95....8K}.
 For the star 11 Com a UVES \citep{2000SPIE.4008..534D} spectra
 was taken from the ESO archive, HARPS-N spectra \citep{2012SPIE.8446E..1VC} were taken from the
 Telescopio Nazionale Galileo archive\footnote{http://ia2.oats.inaf.it/archives/tng}  for the star KOI-415.
 Additional data for six stars  were taken from our own observations
 \citep{2013A&A...554A..84M,2016A&A...588A..98M} four of  them using the Nordic Optical Telescope (2.56 m) with the FIES instrument
 \citep{1999anot.conf...71F}, and two stars using the MERCATOR telescope (1.2 m) with the HERMES
 spectrograph \citep{2011A&A...526A..69R}.
 Table~\ref{tabla_espectrografos} summarises the properties of the different spectra.

\begin{table}
\centering
\caption{
Properties of the different spectrographs used in this work.
}
\label{tabla_espectrografos}
\begin{scriptsize}
\begin{tabular}{lccc}
\hline\noalign{\smallskip}
Spectrograph &  Spectral range (\AA)  & Resolving power  & $N$ stars \\
\hline 
 SOPHIE       &  3872-6943             & 75000            & 20  \\
HARPS        &  3780-6910             & 115000           & 16 \\
ELODIE       &  3850-6800             & 42000            & 6  \\ 
FIES         &  3640-7360             &  67000           & 4  \\
FEROS        &  3500-9200             &  48000           & 3  \\
HERMES       &  3800-9000             &  85000           & 2  \\
HARPS-N      &  3830-6930             & 115000           & 1  \\ 
UVES         &  4780-6800             & 110000           & 1  \\  
\hline
\end{tabular}
\end{scriptsize}
\end{table}

 All the spectra were reduced by the corresponding pipelines which implement the typical corrections 
 involved in \'echelle spectra reduction. 
 When needed several spectra of the same star were properly combined
 in order to obtain a higher signal-to-noise ratio  (S/N) spectra.
 Typical values of the S/N (measured around 605 nm) are between 70
 and 200.
 The spectra were corrected from radial velocity shifts by using
 the precise radial velocities provided by the ELODIE, SOPHIE and
 HARPS data reduction pipelines. For the rest of the targets
 radial velocities were
 measured by cross-correlating their spectra  
 with spectra of radial velocity standard stars of similar
 spectral types obtained during the same observations.

\subsection{Analysis}
 
 Basic stellar parameters (T$_{\rm eff}$, $\log g$, microturbulent
 velocity $\xi_{\rm t}$, and [Fe/H]) were determined by applying the iron ionisation
 and excitation equilibrium conditions to a set of well defined
 302 Fe~{\sc i} and 28 Fe~{\sc ii} lines. The computations
 were done with the {\sc TGVIT}\footnote{http://optik2.mtk.nao.ac.jp/\textasciitilde{}takeda/tgv/}
 \citep{2005PASJ...57...27T} code.
 The line list and details on the adopted parameters
 (excitation potential,
 $\log (gf)$ values) are available on Y. Takeda's web page. 
 ATLAS9, plane-parallel, LTE atmosphere models
 \citep{1993KurCD..13.....K} were used in the computations.
 Uncertainties in the stellar parameters are obtained by
 changing each stellar parameter from the converged solution
 until the excitation equilibrium, the ionisation equilibrium 
 or the match of the curve of growth
 is no longer fulfilled. 
 We are aware that this procedure only evaluates ``statistical'' errors and that
 other systematic sources of uncertainties  
 (i.e., the list lines used, the adopted atomic parameters, or the choice of the atmosphere model)
 are not taken into account \citep[see, for details][]{2002PASJ...54..451T,2002PASJ...54.1041T}.
 \cite{2012A&A...547A..91Z} estimated that more realistic uncertainties
 might be of the order of two-three times the ones provided by TGVIT.  

 In order to avoid weak lines as well as errors due to 
 uncertainties in the damping parameters
 only lines with measured equivalent widths (EWs) between 8 and 120 m\AA \space
 were used
 \citep[e.g.][]{2008PASJ...60..781T}.
 Stellar EWs were measured using the automatic code ARES2
 \citep{2015A&A...577A..67S} 
 adjusting the reject parameter according to the S/N ratio of the spectra as
 described in \cite{2008A&A...487..373S}.

\subsection{Photometric parameters and comparison with previous works}
 
\begin{figure*}[!htb]
\centering
\begin{minipage}{0.33\linewidth}
\includegraphics[angle=270,scale=0.39]{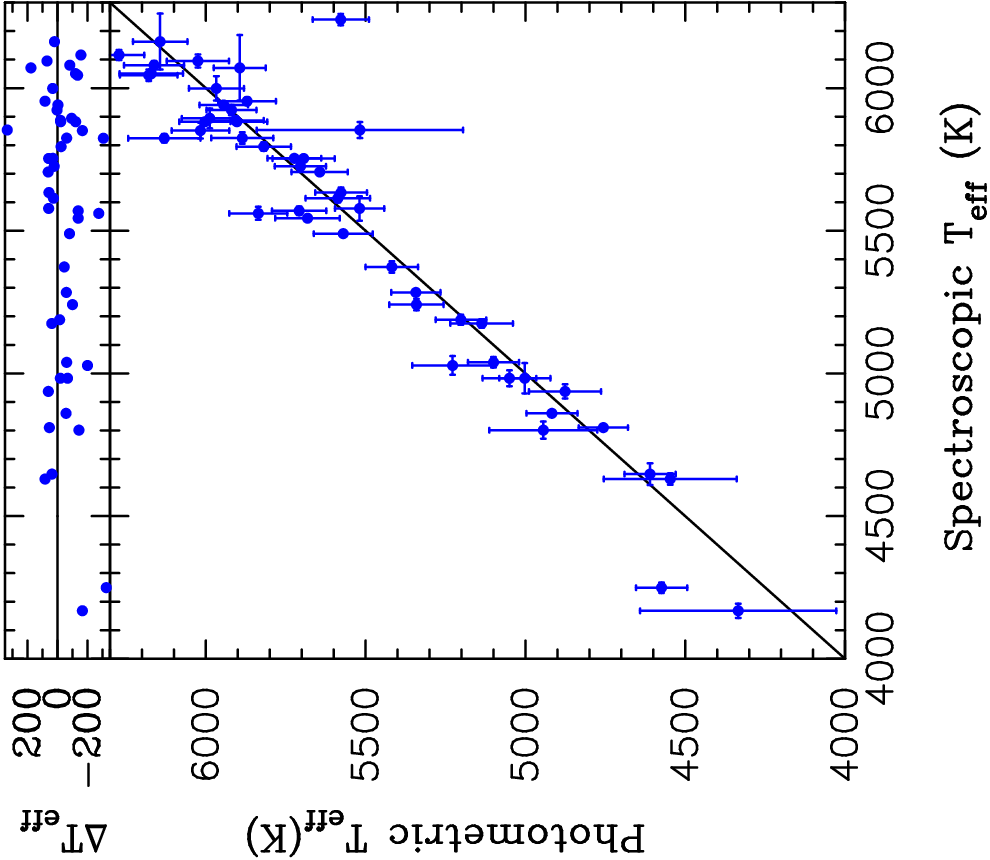}
\end{minipage}
\begin{minipage}{0.33\linewidth}
\includegraphics[angle=270,scale=0.39]{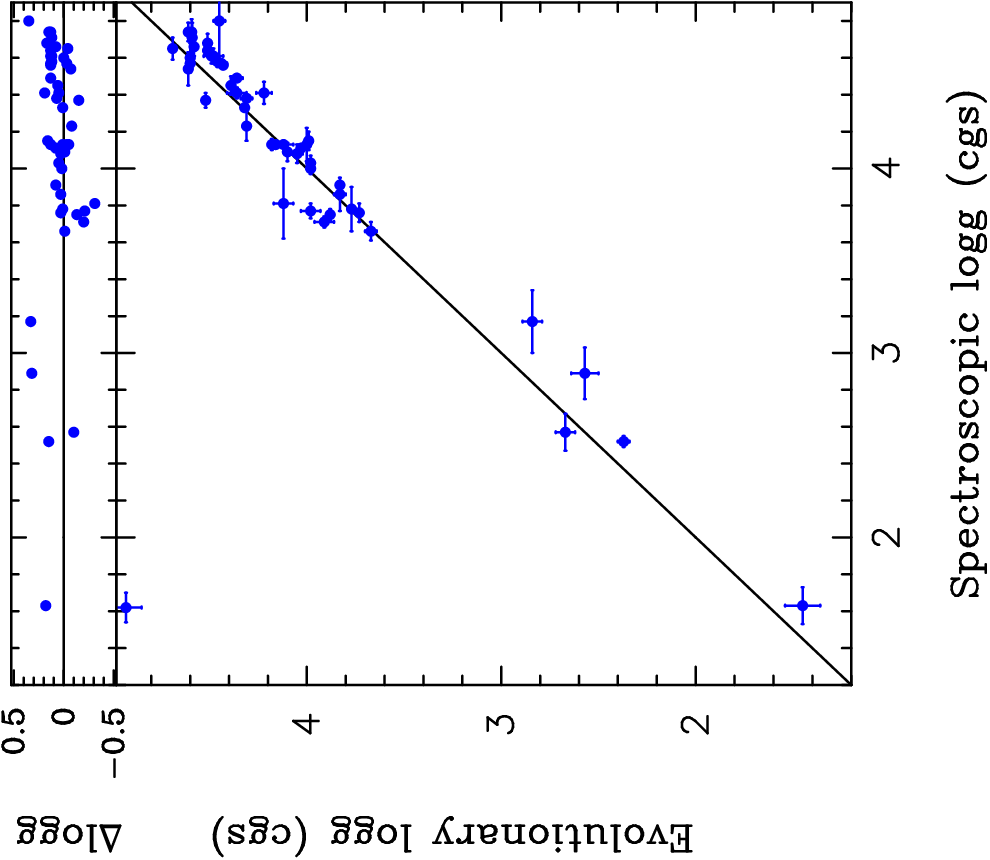}
\end{minipage}
\begin{minipage}{0.33\linewidth}
\includegraphics[angle=270,scale=0.39]{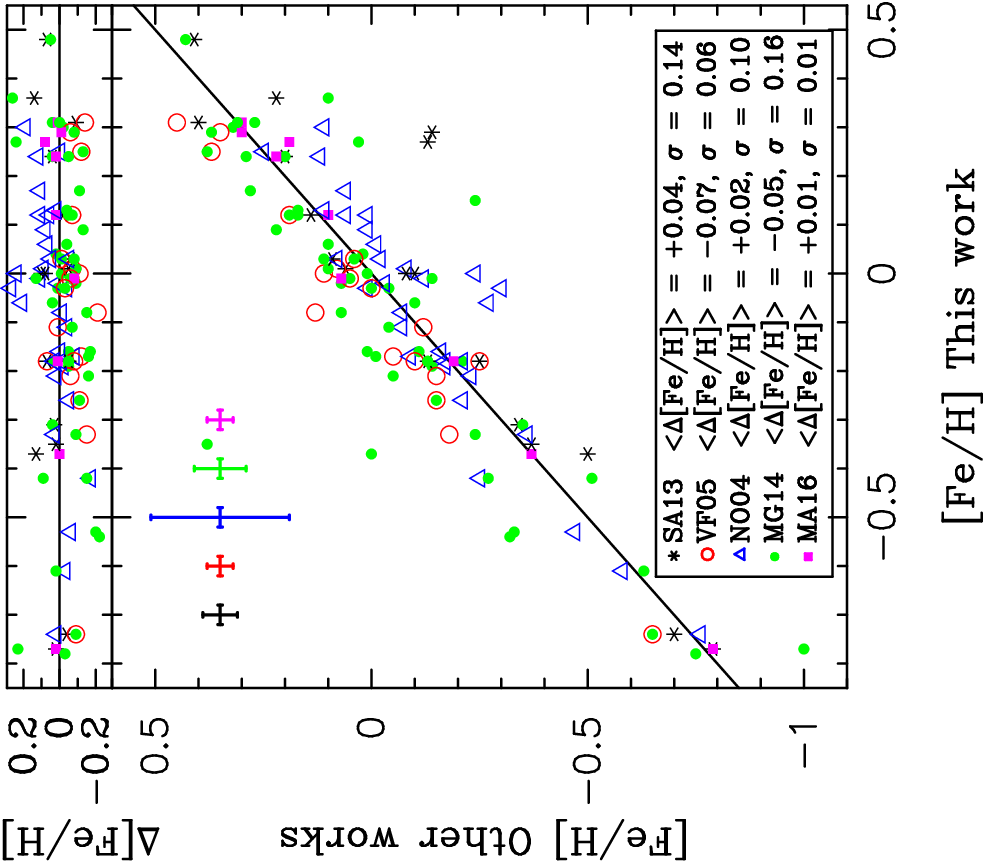}
\end{minipage}
\caption{  Spectroscopic stellar parameters derived in this work
versus estimates based on photometry or values from the literature. 
Left: effective temperature; Middle: surface gravity $\log g$;
Right: stellar metallicity [Fe/H].
The upper
panels show the differences between the parameters derived in this
work and the values obtained from photometry or given in the literature.}
\label{photometric_parameters}
\end{figure*}

 In order to test the reliability of our derived parameters, photometric
 effective temperatures were derived from the {\sc Hipparcos}
 $(B-V)$ colours \citep{1997ESASP1200.....P} by using the calibration provided by 
 \citet[][Table~4]{2010A&A...512A..54C}. Before computation,
 colours were de-reddened by using the stellar galactic
 coordinates and the tables given by \cite{1992A&A...258..104A}.
 Distances were obtained from the revised
 parallaxes provided by \cite{Leeuwen} from a new reduction of the
 {\sc Hipparcos}’s raw data. 
 In the few cases in which colours or parallaxes were not available
 we took the values provided by the Simbad database \citep{2000A&AS..143....9W}.
 The comparison between the temperature values obtained by both
 procedures, spectroscopic and photometric, is shown in 
 the left panel of Figure~\ref{photometric_parameters}. It is clear from the figure
 that there is  no sound systematic difference between them.
 Both temperatures differ in median by 21 K, with a standard
 deviation  of 153 K. 

 Stellar evolutionary parameters, namely surface gravity,
 age, mass, and radius were computed from  
 {\sc Hipparcos} V magnitudes and parallaxes using the code
 {\sc PARAM}\footnote{http://stev.oapd.inaf.it/cgi-bin/param}
 \citep{2006A&A...458..609D}, which is based on the use of
 Bayesian methods, and the {\sc PARSEC} set of isochrones by
 \cite{2012MNRAS.427..127B}.
 The comparison between the spectroscopic and evolutionary
 $\log g$ values is shown in the middle panel
 of Figure~\ref{photometric_parameters}.
 The figure reveals the known trend of
 spectroscopic surface gravities to be systematically
 larger than the 
 evolutionary estimates \citep[e.g.][]{2006A&A...458..609D,2013A&A...554A..84M}.
 Besides that, the distribution of $\log g$$_{\rm spec}$ - $\log g$$_{\rm evol}$ 
 shows a median value of only 0.05, and a standard deviation of 0.13
  consistent with previous works \citep{2006A&A...458..609D,2013A&A...554A..84M,2015A&A...579A..20M}.
 The outlier in  upper left corner is BD+20 2457,
 which has a largely undetermined parallax,
 $\pi$ = 5.0 $\pm$ 26.0 mas \citep{2009ApJ...707..768N}.

 We finally compare our metallicities with those already reported in the literature.
 Values for the comparison are taken from {\it i)} the
 SWEETCat catalogue \citep[][SA13]{2013A&A...556A.150S},
 whose parameters are mainly derived from the same authors using
 the iron ionisation and equilibrium conditions;
 {\it ii)} \citet[][VF05]{2005ApJS..159..141V},  where metallicities are computed by using
 spectral synthesis; {\it iii)} \citet[][NO04]{2004A&A...418..989N}, which provide
 photometric metallicities; {\it iv)} the compilations
 by \citet[][MG14]{2014MNRAS.439.2781M} and \citet{2016A&A...588A.144W};
 and from {\it v)} \citet[][MA16]{2016A&A...588A..98M}
 as a consistency double check. The comparison is shown
 in the right panel of Figure~\ref{photometric_parameters}.
 The agreement is overall good and no systematic differences
 are found with the literature estimates.

\onllongtab{
\begin{longtab}
\begin{landscape}
\begin{longtable}{lccccccccccc} 
\caption{
 Spectroscopic parameters with uncertainties
 for the stars measured in this work. 
  }\label{parameters_table_full}\\
\hline
\hline
 Star &   M$_{\rm C}$$\sin i$$^{\star}$  &  T$_{\rm eff}$ &  $\log g$            & $\xi_{t}$           & [Fe/H] & Sp.$^{\dag}$   &  Age  & M$_{\star}$   & R$_{\star}$ & LC$^{\sharp}$ &  Kin$^{\ddag}$   \\
     & (M$_{\rm Jup}$) &       (K)          & (${\rm cm s^{-2}}$)  & (${\rm km s^{-1}}$) & (dex)    &                & (Gyr) & (M$_{\odot}$) & (R$_{\odot}$) &  & \\
 (1) & (2)&  (3)          & (4)                  & (5)                 & (6)    & (7)            & (8)   & (9)           & (10)     & (11) & (12) \\ 
\hline
\endfirsthead
\caption{Continued.} \\
\hline
 Star &   M$_{\rm C}$$\sin i$$^{\star}$  &  T$_{\rm eff}$ &  $\log g$            & $\xi_{t}$           & [Fe/H] & Sp.$^{\dag}$   &  Age  & M$_{\star}$   & R$_{\star}$ & LC$^{\sharp}$ &  Kin$^{\ddag}$   \\
     & (M$_{\rm Jup}$) &       (K)          & (${\rm cm s^{-2}}$)  & (${\rm km s^{-1}}$) & (dex)    &                & (Gyr) & (M$_{\odot}$) & (R$_{\odot}$) &  & \\
 (1) & (2)&  (3)          & (4)                  & (5)                 & (6)    & (7)            & (8)   & (9)           & (10)     & (11) & (12) \\ 
\hline
\endhead
\hline
\endfoot
\hline
\endlastfoot
\hline\noalign{\smallskip}
HD 4747	&	46.1	$\pm$	2.3	a	&	5373	$\pm$	20	&	4.66	$\pm$	0.05	&	0.84	$\pm$	0.19	&	-0.18	$\pm$	0.02	&	5	&	1.53	$\pm$	1.39	&	0.85	$\pm$	0.01	&	0.76	$\pm$	0.01	&	5	&	D	\\
HD 5388	&	69.2	$\pm$	19.9	a,tm	&	6116	$\pm$	18	&	3.75	$\pm$	0.03	&	1.55	$\pm$	0.08	&	-0.42	$\pm$	0.01	&	2	&	5.29	$\pm$	0.34	&	1.10	$\pm$	0.02	&	1.93	$\pm$	0.07	&	4	&	D	\\
HIP 5158	&	15.04	$\pm$	10.55	a	&	4750	$\pm$	35	&	4.71	$\pm$	0.08	&	0.35	$\pm$	0.39	&	0.36	$\pm$	0.04	&	2	&	3.04	$\pm$	3.17	&	0.80	$\pm$	0.02	&	0.73	$\pm$	0.02	&	5	&	D	\\
HD 10697	&	38	$\pm$	13	a,tm	&	5634	$\pm$	18	&	4.03	$\pm$	0.04	&	1.14	$\pm$	0.16	&	0.12	$\pm$	0.03	&	4	&	7.40	$\pm$	0.22	&	1.11	$\pm$	0.02	&	1.74	$\pm$	0.03	&	4	&	D	\\
HD 13189	&	20			a	&	4168	$\pm$	25	&	1.63	$\pm$	0.10	&	1.57	$\pm$	0.15	&	-0.37	$\pm$	0.05	&	4	&	4.50	$\pm$	2.88	&	1.23	$\pm$	0.25	&	33.69	$\pm$	5.93	&	3	&	D	\\
HD 13507	&	67$^{\rm +8}_{\rm -9}$			b	&	5726	$\pm$	18	&	4.61	$\pm$	0.04	&	0.94	$\pm$	0.10	&	-0.03	$\pm$	0.02	&	1	&	1.57	$\pm$	1.19	&	0.99	$\pm$	0.02	&	0.91	$\pm$	0.01	&	5	&	D	\\
HD 14348	&	48.9	$\pm$	1.6	b	&	6095	$\pm$	23	&	4.09	$\pm$	0.05	&	1.26	$\pm$	0.09	&	0.17	$\pm$	0.02	&	1	&	3.19	$\pm$	0.10	&	1.31	$\pm$	0.02	&	1.63	$\pm$	0.05	&	4	&	D	\\
HD 14651	&	47	$\pm$	3.4	a	&	5490	$\pm$	8	&	4.57	$\pm$	0.02	&	0.83	$\pm$	0.06	&	-0.03	$\pm$	0.01	&	2	&	8.36	$\pm$	2.80	&	0.89	$\pm$	0.02	&	0.91	$\pm$	0.03	&	5	&	TR	\\
HD 16760	&	13.13	$\pm$	0.56	a	&	5614	$\pm$	15	&	4.61	$\pm$	0.04	&	0.69	$\pm$	0.09	&	-0.02	$\pm$	0.01	&	1	&	2.78	$\pm$	2.72	&	0.95	$\pm$	0.02	&	0.91	$\pm$	0.04	&	5	&	D	\\
HD 22781	&	13.65	$\pm$	0.97	a	&	5175	$\pm$	15	&	4.57	$\pm$	0.04	&	0.15	$\pm$	0.35	&	-0.35	$\pm$	0.02	&	1	&	4.14	$\pm$	3.63	&	0.75	$\pm$	0.02	&	0.70	$\pm$	0.02	&	5	&	D	\\
HD 283668	&	53	$\pm$	4	b	&	4860	$\pm$	25	&	4.65	$\pm$	0.06	&	0.03	$\pm$	0.25	&	-0.78	$\pm$	0.01	&	1	&	5.90	$\pm$	4.22	&	0.62	$\pm$	0.01	&	0.57	$\pm$	0.01	&	5	&	TR	\\
HIP 21832	&	40.9	$\pm$	26.2	a,tm	&	5570	$\pm$	15	&	4.37	$\pm$	0.04	&	0.87	$\pm$	0.11	&	-0.61	$\pm$	0.01	&	1	&	11.33	$\pm$	0.30	&	0.74	$\pm$	0.00	&	0.76	$\pm$	0.00	&	5	&	TR	\\
HD 30246	&	55.1$^{\rm +20.3}_{\rm -8.2}$		a	&	5795	$\pm$	15	&	4.58	$\pm$	0.03	&	1.10	$\pm$	0.07	&	0.12	$\pm$	0.01	&	2	&	0.95	$\pm$	0.81	&	1.07	$\pm$	0.01	&	0.98	$\pm$	0.03	&	5	&	D	\\
HD 39091	&	10.27	$\pm$	0.84	b	&	5941	$\pm$	10	&	4.33	$\pm$	0.02	&	1.15	$\pm$	0.04	&	0.03	$\pm$	0.01	&	2	&	4.96	$\pm$	0.26	&	1.07	$\pm$	0.01	&	1.15	$\pm$	0.00	&	5	&	D	\\
HD 38529	&	13.99	$\pm$	0.59	a	&	5578	$\pm$	43	&	3.78	$\pm$	0.12	&	1.18	$\pm$	0.14	&	0.31	$\pm$	0.04	&	6	&	3.88	$\pm$	0.11	&	1.38	$\pm$	0.02	&	2.47	$\pm$	0.07	&	4	&	D	\\
HD 39392	&	13.2	$\pm$	0.8	b	&	5824	$\pm$	15	&	3.71	$\pm$	0.03	&	1.14	$\pm$	0.06	&	-0.54	$\pm$	0.01	&	1	&	9.06	$\pm$	1.40	&	0.94	$\pm$	0.04	&	1.72	$\pm$	0.14	&	4	&	TR	\\
NGC 2423-3	&	10.64	$\pm$	0.93	a	&	4630	$\pm$	20	&	2.44	$\pm$	0.08	&	1.35	$\pm$	0.09	&	0.00	$\pm$	0.04	&	2	&				&				&				&	3	&	D	\\
HD 65430	&	67.8			a	&	5188	$\pm$	18	&	4.68	$\pm$	0.05	&	0.50	$\pm$	0.28	&	-0.11	$\pm$	0.02	&	1	&	10.13	$\pm$	1.51	&	0.80	$\pm$	0.01	&	0.79	$\pm$	0.01	&	5	&	D	\\
HD 72946	&	60.4	$\pm$	2.2	b	&	6240	$\pm$	20	&	4.29	$\pm$	0.05	&	1.35	$\pm$	0.12	&	0.03	$\pm$	0.02	&	3	&				&				&				&	5	&	D	\\
HAT-P-13	&	14.28	$\pm$	0.28	a	&	5853	$\pm$	28	&	4.41	$\pm$	0.06	&	0.91	$\pm$	0.15	&	0.48	$\pm$	0.03	&	1	&	3.02	$\pm$	0.35	&	1.22	$\pm$	0.03	&	1.38	$\pm$	0.08	&	4	&	D	\\
HD 77065	&	41	$\pm$	2	b	&	5039	$\pm$	18	&	4.74	$\pm$	0.05	&	0.07	$\pm$	0.38	&	-0.42	$\pm$	0.02	&	1	&	7.59	$\pm$	3.69	&	0.71	$\pm$	0.01	&	0.67	$\pm$	0.02	&	5	&	TR	\\
BD+26 1888	&	26	$\pm$	2	b	&	4798	$\pm$	40	&	4.54	$\pm$	0.09	&	0.56	$\pm$	0.35	&	0.04	$\pm$	0.04	&	1	&	2.90	$\pm$	3.13	&	0.77	$\pm$	0.02	&	0.70	$\pm$	0.02	&	5	&	D	\\
BD+20 2457	&	22.7	$\pm$	8.1	a	&	4249	$\pm$	18	&	1.62	$\pm$	0.08	&	1.58	$\pm$	0.09	&	-0.77	$\pm$	0.03	&	4	&	4.54	$\pm$	4.06	&	0.40	$\pm$	0.02	&	0.34	$\pm$	0.04	&	3	&	TD	\\
BD+20 2457      &       13.2    $\pm$   4.7     a       &  \multicolumn{10}{c}{} \\
HD 89707	&	53.6$^{\rm +7.8}_{\rm -6.9}$		a	&	5894	$\pm$	35	&	4.23	$\pm$	0.08	&	1.18	$\pm$	0.22	&	-0.53	$\pm$	0.03	&	3	&	11.13	$\pm$	0.49	&	0.84	$\pm$	0.01	&	1.02	$\pm$	0.02	&	5	&	TD	\\
HD 92320	&	59.4	$\pm$	4.1	a	&	5706	$\pm$	10	&	4.64	$\pm$	0.03	&	0.64	$\pm$	0.10	&	-0.06	$\pm$	0.01	&	1	&	0.78	$\pm$	0.70	&	0.98	$\pm$	0.01	&	0.88	$\pm$	0.02	&	5	&	D	\\
11 Com	&	19.4	$\pm$	1.5	a	&	4810	$\pm$	8	&	2.52	$\pm$	0.03	&	1.38	$\pm$	0.07	&	-0.31	$\pm$	0.02	&	8	&	1.17	$\pm$	0.28	&	2.02	$\pm$	0.11	&	14.88	$\pm$	0.36	&	3	&	TR	\\
NGC 4349-127	&	20	$\pm$	1.73	a	&	4439	$\pm$	28	&	1.85	$\pm$	0.11	&	1.58	$\pm$	0.13	&	-0.18	$\pm$	0.05	&	2	&				&				&				&	3	&	TD	\\
HD 114762	&	10.99	$\pm$	0.09	a	&	5851	$\pm$	28	&	4.15	$\pm$	0.05	&	1.25	$\pm$	0.18	&	-0.74	$\pm$	0.02	&	3	&	11.48	$\pm$	0.01	&	0.85	$\pm$	0.00	&	1.52	$\pm$	0.04	&	5	&	TD	\\
HD 122562	&	24	$\pm$	2	b	&	4983	$\pm$	28	&	3.86	$\pm$	0.09	&	0.78	$\pm$	0.16	&	0.31	$\pm$	0.04	&	1	&	7.97	$\pm$	0.97	&	1.12	$\pm$	0.04	&	2.06	$\pm$	0.09	&	4	&	D	\\
HD 132032	&	70	$\pm$	4	b	&	5954	$\pm$	13	&	4.41	$\pm$	0.03	&	0.98	$\pm$	0.08	&	0.09	$\pm$	0.01	&	1	&	2.87	$\pm$	1.54	&	1.10	$\pm$	0.01	&	1.12	$\pm$	0.06	&	5	&	D	\\
HD 131664	&	23$^{\rm +26.0}_{\rm -5.0}$		a,tm	&	5882	$\pm$	8	&	4.49	$\pm$	0.02	&	1.09	$\pm$	0.06	&	0.30	$\pm$	0.01	&	2	&	2.12	$\pm$	1.06	&	1.15	$\pm$	0.01	&	1.14	$\pm$	0.05	&	5	&	D	\\
HD 134113	&	47$^{\rm +2}_{\rm -3}$		b	&	5561	$\pm$	23	&	3.76	$\pm$	0.05	&	1.03	$\pm$	0.10	&	-0.92	$\pm$	0.02	&	1	&	10.98	$\pm$	0.66	&	0.85	$\pm$	0.02	&	2.01	$\pm$	0.07	&	4	&	TD	\\
HD 136118	&	12	$\pm$	0.47	a	&	6163	$\pm$	98	&	3.81	$\pm$	0.19	&	1.32	$\pm$	0.23	&	-0.17	$\pm$	0.06	&	1	&	4.94	$\pm$	1.05	&	1.12	$\pm$	0.04	&	1.48	$\pm$	0.06	&	4	&	D	\\
HD 137759	&	12.7	$\pm$	1.08	a	&	4647	$\pm$	38	&	2.89	$\pm$	0.14	&	1.16	$\pm$	0.21	&	0.27	$\pm$	0.07	&	6	&	2.07	$\pm$	0.74	&	1.78	$\pm$	0.23	&	11.14	$\pm$	0.34	&	3	&	D	\\
HD 137510	&	27.3	$\pm$	1.9	a	&	5999	$\pm$	43	&	4.13	$\pm$	0.09	&	1.31	$\pm$	0.12	&	0.25	$\pm$	0.03	&	1	&	3.15	$\pm$	0.31	&	1.39	$\pm$	0.03	&	1.89	$\pm$	0.06	&	4	&	D	\\
HD 140913	&	43.2	a			&	6071	$\pm$	115	&	4.80	$\pm$	0.24	&	1.68	$\pm$	0.50	&	-0.08	$\pm$	0.07	&	1	&	1.62	$\pm$	1.58	&	1.02	$\pm$	0.04	&	0.96	$\pm$	0.03	&	5	&	D	\\
HD 156846	&	10.57	$\pm$	0.29	b	&	6051	$\pm$	13	&	4.00	$\pm$	0.03	&	1.39	$\pm$	0.05	&	0.13	$\pm$	0.01	&	2	&	3.38	$\pm$	0.41	&	1.36	$\pm$	0.04	&	1.91	$\pm$	0.05	&	4	&	D	\\
HD 160508	&	48	$\pm$	3	b	&	6045	$\pm$	20	&	3.77	$\pm$	0.04	&	1.33	$\pm$	0.08	&	-0.16	$\pm$	0.02	&	1	&	5.55	$\pm$	0.57	&	1.14	$\pm$	0.04	&	1.76	$\pm$	0.13	&	4	&	D	\\
HD 162020	&	14.4	$\pm$	0.04	a	&	4801	$\pm$	30	&	4.60	$\pm$	0.08	&	0.62	$\pm$	0.34	&	0.00	$\pm$	0.04	&	2	&	3.32	$\pm$	3.35	&	0.76	$\pm$	0.02	&	0.70	$\pm$	0.02	&	5	&	D	\\
HD 167665	&	50.6	$\pm$	1.7	a	&	6080	$\pm$	15	&	4.13	$\pm$	0.03	&	1.29	$\pm$	0.08	&	-0.21	$\pm$	0.01	&	5	&	6.72	$\pm$	0.23	&	1.03	$\pm$	0.01	&	1.32	$\pm$	0.02	&	5	&	D	\\
HD 168443	&	34.3	$\pm$	9	a,tm	&	5544	$\pm$	5	&	4.11	$\pm$	0.02	&	1.04	$\pm$	0.04	&	0.01	$\pm$	0.01	&	2	&	10.70	$\pm$	0.46	&	1.00	$\pm$	0.01	&	1.54	$\pm$	0.04	&	4	&	D	\\
HD 174457	&	65.8			b	&	5825	$\pm$	20	&	4.08	$\pm$	0.05	&	1.22	$\pm$	0.12	&	-0.26	$\pm$	0.02	&	3	&	9.80	$\pm$	0.55	&	0.96	$\pm$	0.02	&	1.48	$\pm$	0.07	&	4	&	D	\\
HD 175679	&	37.3	$\pm$	2.8	a	&	5028	$\pm$	33	&	2.57	$\pm$	0.10	&	1.43	$\pm$	0.17	&	-0.01	$\pm$	0.05	&	5	&	0.66	$\pm$	0.11	&	2.53	$\pm$	0.12	&	11.79	$\pm$	0.80	&	3	&	D	\\
HD 180314	&	22			a	&	4983	$\pm$	53	&	3.17	$\pm$	0.17	&	1.27	$\pm$	0.21	&	0.24	$\pm$	0.07	&	4	&	1.14	$\pm$	0.24	&	2.13	$\pm$	0.13	&	8.88	$\pm$	0.47	&	3	&	TR	\\
KOI-415	&	62.14	$\pm$	2.69	b	&	5513	$\pm$	78	&	4.36	$\pm$	0.20	&	0.80	$\pm$	0.31	&	0.15	$\pm$	0.07	&	7	&				&				&				&	4	&		\\
HD 190228	&	49.4	$\pm$	14.8	a,tm	&	5241	$\pm$	20	&	3.66	$\pm$	0.05	&	0.96	$\pm$	0.09	&	-0.33	$\pm$	0.02	&	3	&	5.70	$\pm$	0.69	&	1.12	$\pm$	0.04	&	2.48	$\pm$	0.11	&	4	&	D	\\
HR 7672	&	68.7	$\pm$	3	a,tm	&	5923	$\pm$	18	&	4.45	$\pm$	0.05	&	1.14	$\pm$	0.12	&	-0.01	$\pm$	0.02	&	3	&	3.68	$\pm$	0.71	&	1.05	$\pm$	0.02	&	1.05	$\pm$	0.01	&	5	&	D	\\
HD 191760	&	38.17	$\pm$	1.02	a	&	5887	$\pm$	10	&	4.13	$\pm$	0.02	&	1.22	$\pm$	0.05	&	0.24	$\pm$	0.01	&	2	&	4.33	$\pm$	0.57	&	1.23	$\pm$	0.04	&	1.55	$\pm$	0.11	&	4	&	D	\\
HD 202206	&	17.5			a	&	5754	$\pm$	8	&	4.56	$\pm$	0.02	&	1.03	$\pm$	0.05	&	0.29	$\pm$	0.01	&	2	&	1.02	$\pm$	0.83	&	1.10	$\pm$	0.01	&	1.02	$\pm$	0.03	&	5	&	D	\\
HD 209262	&	32.3$^{\rm +1.6}_{\rm -1.5}$	b	&	5753	$\pm$	8	&	4.38	$\pm$	0.02	&	1.02	$\pm$	0.04	&	0.06	$\pm$	0.01	&	2	&	7.48	$\pm$	0.75	&	1.00	$\pm$	0.01	&	1.13	$\pm$	0.05	&	5	&	D	\\
BD+24 4697	&	53	$\pm$	3	b	&	4937	$\pm$	25	&	4.74	$\pm$	0.07	&	0.12	$\pm$	0.46	&	-0.16	$\pm$	0.03	&	1	&	5.207	$\pm$	4.15	&	0.754	$\pm$	0.016	&	0.705	$\pm$	0.017	&	5	&	TR	\\
HD 217786	&	13	$\pm$	0.8	a	&	5882	$\pm$	8	&	4.13	$\pm$	0.02	&	1.18	$\pm$	0.05	&	-0.19	$\pm$	0.01	&	2	&	9.40	$\pm$	0.22	&	0.96	$\pm$	0.01	&	1.32	$\pm$	0.06	&	4	&	D	\\
HD 219077	&	10.39	$\pm$	0.09	b	&	5284	$\pm$	5	&	3.91	$\pm$	0.02	&	0.94	$\pm$	0.04	&	-0.18	$\pm$	0.01	&	2	&	8.55	$\pm$	0.25	&	1.03	$\pm$	0.01	&	1.99	$\pm$	0.02	&	4	&	TR	\\
\hline\noalign{\smallskip}
\end{longtable}
\tablefoot{
 $^{\star}$ ({\bf a}) \citet[][and references therein]{2014MNRAS.439.2781M}, ({\bf b}) \citet[][and references therein]{2016A&A...588A.144W},
 ({\bf tm}) ``true'' mass.\\
 $^{\dag}$Spectrograph: {\bf(1)} SOPHIE; {\bf(2)} ESO/HARPS; {\bf(3)} ELODIE;
 {\bf(4)} NOT/FIES;
 {\bf(5)} ESO/FEROS; {\bf(6)} MERCATOR/HERMES; {\bf(7)} TNG/HARPS-N; {\bf(8)} ESO/UVES.\\
 $^{\sharp}$ {\bf 5}: Main-sequence, {\bf 4}: Subgiant, {\bf 3}: Giant.\\
$^{\ddag}$ {\bf D}: Thin disc, {\bf TD}: Thick disc, {\bf TR}: Transition.
}%
\end{landscape}
\end{longtab}
}

\subsection{Abundance computation}
 Chemical abundance of individual elements
 C, O, Na, Mg, Al, Si, S, Ca, Sc, Ti, V, Cr,  Mn, Co, Ni,
 and Zn were obtained using the 2014 version of the
 code {\sc MOOG}\footnote{http://www.as.utexas.edu/\textasciitilde{}chris/moog.html}
 \citep{1973PhDT.......180S}. The selected
 lines are taken from the list provided by \cite{2015A&A...579A..20M},
 with the only exception of carbon, for which we use
 the lines at 505.2 and 538.0 nm. For Zn abundances,
 the lines at 481.05 and 636.23 nm were considered.

 Hyperfine structure (HFS) was taken into account for
 V~{\sc i}, and Co~{\sc i} abundances.
 HFS corrections for Mn~{\sc i} were not taken into account
 as \cite{2015A&A...579A..20M} found slightly different abundances when considering
 different lines.
 Finally, the oxygen abundance was determined from the forbidden
 [O~{\sc i}] line at 6300\AA \space. Since this line is blended
 with a closer Ni~{\sc i} line \citep[e.g.][]{2001ApJ...556L..63A},
 we first used the MOOG task ewfind to determine the EW of the
 Ni line. This EW was subtracted from the Ni~{\sc i} plus [O~{\sc i}] feature's
 EW. The remaining EW was used for determining the oxygen abundance
 \citep[e.g.][]{2010ApJ...725.2349D}.
 
 We have used three representative stars, namely HD 180314 
 (T$_{\rm eff}$ = 4983 K), HD 38529 (5578 K), and
 HD 191760 (5887 K)
 in order to provide an estimate on how the uncertainties 
 in the atmospheric parameters propagate into the abundance
 calculation. Abundances for each of these three stars
 were recomputed using T$_{\rm eff}$ = T$_{\rm eff}$ +  $\Delta$T$_{\rm eff}$,
 T$_{\rm eff}$ - $\Delta$T$_{\rm eff}$, and similarly for
 $\log g$, $\xi_{\rm t}$, and [Fe/H]. Results are given
 in Table~\ref{abundance_sensitivity}.

\begin{table*}
\centering
\caption{
Abundance sensitivities.
}
\label{abundance_sensitivity}
\begin{tabular}{lrrrrrrrrrrrr}
\hline\noalign{\smallskip}
      &  \multicolumn{4}{c}{HD 180314}    & \multicolumn{4}{c}{HD 38529}   &\multicolumn{4}{c}{HD 191760}   \\
Ion   &  \multicolumn{4}{c}{\hrulefill}   & \multicolumn{4}{c}{\hrulefill} & \multicolumn{4}{c}{\hrulefill}  \\
      & $\Delta$T$_{\rm eff}$ &  $\Delta$$\log g$  &  $\Delta$[Fe/H] & $\Delta$$\xi_{t}$ & $\Delta$T$_{\rm eff}$ &  $\Delta$$\log g$  &  $\Delta$[Fe/H] & $\Delta$$\xi_{t}$ & $\Delta$T$_{\rm eff}$ &  $\Delta$$\log g$ &  $\Delta$[Fe/H] & $\Delta$$\xi_{t}$ \\ 
      & $\pm$53 &  $\pm$0.17 &  $\pm$0.07  & $\pm$0.21 & $\pm$43  & $\pm$0.12 & $\pm$0.04 & $\pm$0.14 & $\pm$10 & $\pm$0.02 & $\pm$0.01 & $\pm$0.05   \\
      &  (K)       &  (cms$^{\rm -2}$)  &  (dex) &  (kms$^{\rm -1}$)   & (K)  &  (cms$^{\rm -2}$)  &  (dex) & (kms$^{\rm -1}$) & (K)  &  (cms$^{\rm -2}$)  &  (dex) & (kms$^{\rm -1}$)   \\
\hline\noalign{\smallskip}
C~{\sc i}      &       0.06    &       0.07    &       $<$ 0.01        &       0.01    &       0.03    &       0.04    &       $<$ 0.01        &       0.01    &       0.01    &       0.01    &       $<$ 0.01        &       $<$ 0.01        \\      
O~{\sc i}     &       0.01    &       0.07    &       0.03    &       0.02    &       $<$ 0.01        &       0.06    &       0.01    &       0.02    &       $<$ 0.01        &       0.01    &       $<$ 0.01        &       $<$ 0.01        \\      
Na~{\sc i}     &       0.04    &       0.02    &       $<$ 0.01        &       0.08    &       0.02    &       0.01    &       $<$ 0.01        &       0.02    &       0.01    &       $<$ 0.01        &       $<$ 0.01        &       0.01    \\      
Mg~{\sc i}     &       0.03    &       0.02    &       0.01    &       0.08    &       0.03    &       0.02    &       $<$ 0.01        &       0.04    &       0.01    &       $<$ 0.01        &       $<$ 0.01        &       0.01    \\      
Al~{\sc i}     &       0.03    &       0.01    &       $<$ 0.01        &       0.05    &       0.02    &       0.01    &       $<$ 0.01        &       0.02    &       0.01    &       $<$ 0.01        &       $<$ 0.01        &       $<$ 0.01        \\
Si~{\sc i}     &       0.02    &       0.03    &       0.01    &       0.04    &       $<$ 0.01        &       0.01    &       $<$ 0.01        &       0.02    &       $<$ 0.01        &       $<$ 0.01        &       $<$ 0.01        &       0.01    \\      
S~{\sc i}      &       0.06    &       0.06    &       $<$ 0.01        &       0.02    &       0.03    &       0.03    &       $<$ 0.01        &       0.01    &       0.01    &       $<$ 0.01        &       $<$ 0.01        &       $<$ 0.01        \\      
Ca~{\sc i}     &       0.05    &       0.03    &       $<$ 0.01        &       0.10    &       0.03    &       0.02    &       $<$ 0.01        &       0.05    &       0.01    &       $<$ 0.01        &       $<$ 0.01        &       0.01    \\      
Sc~{\sc i}     &       0.07    &       $<$ 0.01        &       $<$ 0.01        &       0.06    &       0.04    &       $<$ 0.01        &       $<$ 0.01        &       0.01    &       0.01    &       $<$ 0.01        &       $<$ 0.01        &       $<$ 0.01        \\
Sc~{\sc ii}    &       $<$ 0.01        &       0.07    &       0.02    &       0.10    &       $<$ 0.01        &       0.05    &       0.01    &       0.05    &       $<$ 0.01        &       0.01    &       $<$ 0.01        &       0.02    \\      
Ti~{\sc i}     &       0.07    &       0.01    &       $<$ 0.01        &       0.11    &       0.05    &       0.01    &       $<$ 0.01        &       0.04    &       0.01    &       $<$ 0.01        &       $<$ 0.01        &       0.01    \\      
Ti~{\sc ii}    &       0.01    &       0.07    &       0.02    &       0.11    &       $<$ 0.01        &       0.05    &       0.01    &       0.06    &       $<$ 0.01        &       0.01    &       $<$ 0.01        &       0.02    \\      
V~{\sc i}   &       0.09    &       $<$ 0.01        &       0.01    &       0.03    &       0.05    &       $<$ 0.01        &       $<$ 0.01        &       $<$ 0.01        &       0.02    &       $<$ 0.01        &       $<$ 0.01        &       $<$ 0.01        \\
Cr~{\sc i}     &       0.05    &       0.01    &       $<$ 0.01        &       0.09    &       0.03    &       0.01    &       $<$ 0.01        &       0.04    &       0.01    &       $<$ 0.01        &       $<$ 0.01        &       0.01    \\      
Cr~{\sc ii}    &       0.03    &       0.07    &       0.02    &       0.08    &       0.01    &       0.04    &       0.01    &       0.06    &       $<$ 0.01        &       0.01    &       $<$ 0.01        &       0.02    \\      
Mn~{\sc i}     &       0.04    &       0.02    &       0.01    &       0.14    &       0.03    &       0.01    &       $<$ 0.01        &       0.07    &       0.01    &       $<$ 0.01        &       $<$ 0.01        &       0.02    \\      
Co~{\sc i}  &       0.03    &       0.04    &       0.02    &       0.01    &       0.03    &       0.01    &       $<$ 0.01        &       $<$ 0.01        &       0.01    &       $<$ 0.01        &       $<$ 0.01        &       $<$ 0.01        \\
Ni~{\sc i}     &       0.01    &       0.03    &       0.01    &       0.08    &       0.02    &       $<$ 0.01        &       $<$ 0.01        &       0.04    &       0.01    &       $<$ 0.01        &       $<$ 0.01        &       0.01    \\      
Zn~{\sc i}     &       0.03    &       0.05    &       0.02    &       0.11    &       0.01    &       0.02    &       0.01    &       0.06    &       $<$ 0.01        &       $<$ 0.01        &       $<$ 0.01        &       0.02    \\ 
\hline\noalign{\smallskip}
\end{tabular}
\end{table*}

 As final uncertainties for the derived abundances, we give
 the quadratic sum of the uncertainties due to the 
 propagation of the errors in the stellar parameters,
 plus the line-to-line scatter errors.
 For abundances derived from one single line a line-to-line
 scatter error of 0.03 dex (the median value of all
 the scatter errors) was assumed. Abundances with large line-to-line
 scatter errors were discarded. 
 We should caution that these uncertainties should
 be regarded as lower limits given that abundance estimates are
 affected by systematics (i.e. atmosphere models or atomic data)
 that are difficult to account for.
 Our abundances are given in Table~\ref{abundance_table_full}.
 They are expressed relative to the solar values derived
 in \cite{2015A&A...579A..20M} and \cite{2016A&A...588A..98M}  which were obtained by using similar
 spectra and the same methodology
 to the one used in this work. 

\begin{figure*}[!htb]
\centering
\includegraphics[angle=270,scale=0.75]{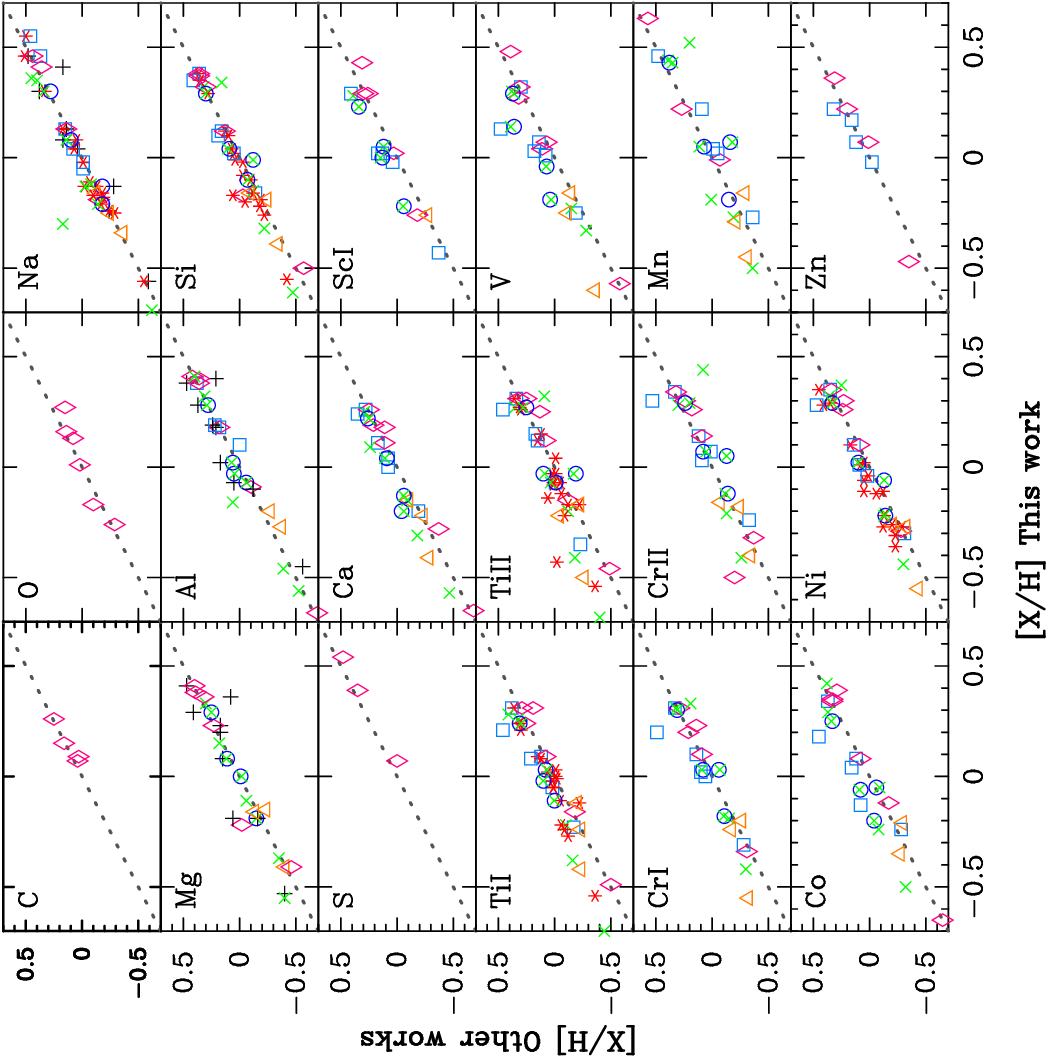}
\caption{
 Comparison of our abundances to those of
 \cite{2005A&A...438..251B} (black $+$), 
 \cite{2005ApJS..159..141V} (red $\ast$),
 \cite{2007PASJ...59..335T} (cyan open squares),
 \cite{2009A&A...497..563N} (blue open circles),
 \cite{2012A&A...545A..32A} (green $\times$),
 \cite{2014A&A...566A..83M} (orange open triangles)
 and \cite{2016A&A...588A..98M} (purple open diamonds).
}
\label{comparison_abundances}
\end{figure*}

 A comparison of our derived abundances with those previously reported in the literature is
 shown in Figure~\ref{comparison_abundances}. Whilst individual
 comparisons among this work and those in the literature are difficult to perform given
 the small number of stars in common 
 and the different species
 analysed, there seems to be an overall good agreement between our estimates
 and previously reported values for the refractory elements.
 In the case of volatile elements (C, O, S, and Zn) we found few previous estimates to compare with,
 most likely due to the inherent difficulties in obtaining accurate abundances for these elements.

\onllongtab{
\begin{longtab}
\begin{landscape}
\begin{longtable}{lrrrrrrrrrrrrrrrrrrr}
\caption{\label{abundance_table_full} Derived abundances [X/H]}\\
\hline\hline
 Star    &          C~{\sc i}  & O~{\sc i}  & Na~{\sc i} & Mg~{\sc i} & Al~{\sc i} &
                   Si~{\sc i} & S~{\sc i}  & Ca~{\sc i} & Sc~{\sc i} & Sc~{\sc ii} &
                   Ti~{\sc i} & Ti~{\sc ii} & V~{\sc i} & Cr~{\sc i} & Cr~{\sc ii} &
                   Mn~{\sc i} & Co~{\sc i}  & Ni~{\sc i} & Zn~{\sc i}\\
\hline
\endfirsthead
\caption{continued.}\\
\hline\hline
 Star   &          C~{\sc i}  & O~{\sc i}  & Na~{\sc i} & Mg~{\sc i} & Al~{\sc i} &
                   Si~{\sc i} & S~{\sc i}  & Ca~{\sc i} & Sc~{\sc i} & Sc~{\sc ii} &
                   Ti~{\sc i} & Ti~{\sc ii} & V~{\sc i} & Cr~{\sc i} & Cr~{\sc ii} &
                   Mn~{\sc i} & Co~{\sc i}  & Ni~{\sc i} & Zn~{\sc i}\\
\hline
\endhead
\hline
\endfoot
\hline\noalign{\smallskip}
HD 4747 &                &         &  -0.25 &   -0.15 &   -0.20 &   -0.19 &   -0.07 &   -0.22 &   -0.26 &   -0.27 &   -0.12 &   -0.17 &   -0.16 &   -0.20 &   -0.18 &   -0.16 &   -0.21 &   -0.27 &   -0.20 \\ 
              &              &                &    0.04 &           0.06 &           0.04 &           0.02 &           0.05 &           0.06 &           0.05 &           0.07 &           0.06 &           0.07 &           0.05 &           0.05 &           0.07 &           0.08 &           0.06 &           0.04 &           0.16 \\         
HD 5388 &         -0.31 &   -0.42 &   -0.30 &   -0.37 &   -0.46 &   -0.32 &          &  -0.31 &          &  -0.51 &   -0.38 &   -0.41 &   -0.33 &   -0.42 &   -0.41 &   -0.50 &   -0.50 &   -0.44 &   -0.53 \\ 
              &  0.03 &           0.04 &           0.06 &           0.07 &           0.03 &           0.02 &                       &    0.04 &                       &    0.04 &           0.03 &           0.04 &           0.06 &           0.03 &           0.06 &           0.06 &           0.04 &           0.02 &           0.05 \\         
HIP 5158 &        1.09 &    -0.05 &   0.36 &    0.15 &    0.41 &    0.34 &           &  0.09 &           &  0.39 &    0.28 &    0.32 &    0.64 &    0.33 &    0.44 &    0.52 &    0.42 &    0.37 &        \\   
              &  0.07 &           0.07 &           0.12 &           0.07 &           0.06 &           0.04 &                       &    0.13 &                       &    0.10 &           0.10 &           0.10 &           0.07 &           0.07 &           0.12 &           0.13 &           0.06 &           0.06 &                       \\  
HD 10697 &        0.07 &    0.01 &    0.13 &    0.23 &    0.18 &    0.12 &    0.07 &    0.11 &    0.02 &    0.11 &    0.09 &    0.12 &    0.07 &    0.10 &    0.14 &    0.22 &    0.08 &    0.10 &    0.07 \\  
              &  0.04 &           0.06 &           0.04 &           0.05 &           0.04 &           0.02 &           0.15 &           0.06 &           0.06 &           0.06 &           0.05 &           0.07 &           0.05 &           0.05 &           0.08 &           0.07 &           0.06 &           0.04 &           0.08 \\         
HD 13189 &        0.15 &    -0.17 &   -0.19 &   -0.22 &   -0.09 &   -0.17 &          &  -0.28 &          &  -0.31 &   -0.16 &   -0.15 &   0.04 &    -0.34 &   -0.32 &   -0.01 &   -0.12 &   -0.29 &       \\   
              &  0.08 &           0.07 &           0.11 &           0.08 &           0.06 &           0.06 &                       &    0.09 &                       &    0.10 &           0.11 &           0.14 &           0.17 &           0.09 &           0.14 &           0.13 &           0.10 &           0.07 &                       \\  
HD 13507 &        -0.13 &   -0.07 &   -0.17 &   -0.09 &   -0.14 &   -0.08 &   0.01 &    -0.04 &   -0.07 &   -0.15 &   -0.05 &   -0.08 &   -0.05 &   -0.02 &   -0.02 &   -0.13 &   -0.14 &   -0.12 &   -0.28 \\ 
              &  0.07 &           0.04 &           0.03 &           0.03 &           0.05 &           0.02 &           0.04 &           0.04 &           0.03 &           0.05 &           0.03 &           0.05 &           0.06 &           0.03 &           0.05 &           0.06 &           0.05 &           0.03 &           0.04 \\         
HD 14348 &        0.14 &    0.09 &    0.29 &    0.16 &    0.14 &    0.20 &    0.05 &    0.18 &    0.16 &    0.14 &    0.13 &    0.12 &    0.16 &    0.13 &    0.20 &    0.22 &    0.06 &    0.17 &    0.07 \\  
              &  0.03 &           0.04 &           0.06 &           0.05 &           0.02 &           0.02 &           0.03 &           0.04 &           0.03 &           0.04 &           0.02 &           0.04 &           0.04 &           0.02 &           0.07 &           0.04 &           0.04 &           0.02 &           0.06 \\         
HD 14651 &        -0.06 &   0.08 &    -0.08 &   0.04 &    0.03 &    -0.03 &   0.02 &    -0.08 &   -0.07 &   -0.06 &   -0.02 &   -0.01 &   -0.02 &   -0.02 &   -0.02 &   0.04 &    -0.03 &   -0.05 &   -0.06 \\ 
              &  0.07 &           0.05 &           0.05 &           0.06 &           0.03 &           0.02 &           0.05 &           0.06 &           0.05 &           0.06 &           0.05 &           0.07 &           0.05 &           0.05 &           0.08 &           0.07 &           0.06 &           0.04 &           0.08 \\         
HD 16760 &              &         &  -0.18 &   -0.14 &   -0.08 &   -0.08 &   -0.19 &   -0.06 &   -0.14 &   -0.15 &   -0.06 &   -0.05 &   -0.07 &   0.00 &    0.04 &    -0.02 &   -0.17 &   -0.09 &   -0.05 \\ 
              &              &                &    0.04 &           0.05 &           0.04 &           0.02 &           0.05 &           0.06 &           0.03 &           0.05 &           0.04 &           0.06 &           0.04 &           0.04 &           0.06 &           0.06 &           0.07 &           0.03 &           0.14 \\         
HD 22781 &        -0.05 &          &  -0.24 &   -0.26 &   0.11 &    -0.28 &          &  -0.21 &   -0.07 &   -0.12 &   -0.05 &   -0.15 &   -0.06 &   -0.27 &   -0.25 &   -0.42 &   -0.25 &   -0.36 &   -0.18 \\ 
              &  0.18 &                       &    0.10 &           0.06 &           0.15 &           0.03 &                       &    0.08 &           0.10 &           0.16 &           0.08 &           0.09 &           0.07 &           0.06 &           0.08 &           0.09 &           0.05 &           0.05 &           0.16 \\         
HD 283668 &       0.45 &           &  -0.75 &   -0.56 &   -0.40 &   -0.57 &          &  -0.53 &   -0.44 &   -0.79 &   -0.41 &   -0.51 &   -0.43 &   -0.66 &   -0.67 &   -0.95 &   -0.63 &   -0.75 &   -0.68 \\ 
              &  0.07 &                       &    0.10 &           0.07 &           0.06 &           0.04 &                       &    0.08 &           0.07 &           0.09 &           0.09 &           0.10 &           0.08 &           0.08 &           0.09 &           0.11 &           0.05 &           0.06 &           0.09 \\         
HIP 21832 &       -0.48 &          &  -0.64 &   -0.34 &   -0.44 &   -0.46 &          &  -0.48 &          &  -0.58 &   -0.43 &   -0.43 &   -0.44 &   -0.62 &   -0.52 &   -0.79 &   -0.54 &   -0.62 &   -0.61 \\ 
              &  0.05 &                       &    0.05 &           0.09 &           0.04 &           0.02 &                       &    0.06 &                       &    0.06 &           0.05 &           0.07 &           0.09 &           0.05 &           0.07 &           0.08 &           0.04 &           0.04 &           0.07 \\         
HD 30246 &        0.01 &           &  0.08 &    0.09 &    0.09 &    0.08 &           &  0.10 &    0.11 &    0.03 &    0.07 &    0.09 &    0.05 &    0.13 &    0.18 &    0.19 &    -0.02 &   0.08 &    0.02 \\  
              &  0.05 &                       &    0.03 &           0.03 &           0.03 &           0.02 &                       &    0.04 &           0.04 &           0.04 &           0.03 &           0.04 &           0.04 &           0.03 &           0.04 &           0.06 &           0.09 &           0.02 &           0.11 \\         
HD 39091 &        -0.03 &   -0.01 &   0.08 &    0.08 &    0.02 &    0.04 &    0.05 &    0.04 &    0.00 &    -0.04 &   -0.02 &   -0.03 &   -0.04 &   0.03 &    0.07 &    0.05 &    -0.06 &   0.02 &    -0.04 \\ 
              &  0.06 &           0.03 &           0.02 &           0.03 &           0.02 &           0.02 &           0.16 &           0.03 &           0.03 &           0.03 &           0.02 &           0.03 &           0.03 &           0.02 &           0.03 &           0.03 &           0.02 &           0.02 &           0.07 \\         
HD 38529 &        0.26 &    0.16 &    0.46 &    0.41 &    0.38 &    0.38 &    0.39 &    0.26 &    0.29 &    0.34 &    0.31 &    0.31 &    0.32 &    0.31 &    0.34 &    0.63 &    0.34 &    0.35 &    0.22 \\  
              &  0.05 &           0.07 &           0.06 &           0.09 &           0.05 &           0.03 &           0.09 &           0.06 &           0.04 &           0.08 &           0.06 &           0.09 &           0.06 &           0.06 &           0.08 &           0.09 &           0.06 &           0.04 &           0.11 \\         
HD 39392 &        -0.38 &          &  -0.36 &   -0.45 &   -0.48 &   -0.44 &   -0.28 &   -0.41 &          &  -0.64 &   -0.52 &   -0.54 &   -0.55 &   -0.56 &   -0.49 &   -0.62 &   -0.41 &   -0.56 &   -0.60 \\ 
              &  0.04 &                       &    0.11 &           0.06 &           0.04 &           0.02 &           0.04 &           0.05 &                       &    0.05 &           0.04 &           0.05 &           0.10 &           0.05 &           0.06 &           0.07 &           0.12 &           0.03 &           0.04 \\         
NGC 2423-3 &      -0.10 &   0.03 &    0.10 &    0.26 &    0.10 &    0.09 &           &  -0.04 &   -0.01 &   -0.02 &   0.05 &    -0.17 &   0.07 &    0.01 &    0.02 &    0.32 &    0.04 &    -0.02 &   -0.43 \\ 
              &  0.05 &           0.06 &           0.09 &           0.06 &           0.04 &           0.04 &                       &    0.08 &           0.21 &           0.08 &           0.08 &           0.14 &           0.06 &           0.06 &           0.08 &           0.10 &           0.05 &           0.05 &           0.08 \\         
HD 65430 &        -0.02 &   -0.05 &   -0.11 &   -0.01 &   0.05 &    -0.02 &          &  -0.16 &   -0.05 &   -0.05 &   0.00 &    0.04 &    -0.01 &   -0.13 &   -0.05 &   -0.09 &   -0.05 &   -0.11 &   0.26 \\  
              &  0.06 &           0.06 &           0.05 &           0.07 &           0.04 &           0.03 &                       &    0.07 &           0.08 &           0.08 &           0.07 &           0.09 &           0.06 &           0.07 &           0.09 &           0.09 &           0.05 &           0.05 &           0.08 \\         
HD 72946 &        0.03 &           &  0.07 &    -0.06 &   0.10 &    0.04 &    -0.05 &   0.01 &           &  -0.13 &   0.00 &    -0.07 &   0.02 &    0.04 &    0.14 &    0.01 &    -0.25 &   -0.01 &   -0.18 \\ 
              &  0.03 &                       &    0.06 &           0.06 &           0.04 &           0.02 &           0.17 &           0.04 &                       &    0.04 &           0.03 &           0.05 &           0.07 &           0.03 &           0.06 &           0.06 &           0.04 &           0.03 &           0.06 \\         
HAT-P-13 &              &         &  0.58 &    0.38 &    0.25 &    0.45 &           &  0.41 &    0.53 &    0.56 &    0.50 &    0.64 &    0.48 &    0.45 &    0.43 &    0.67 &    0.48 &    0.53 &    0.29 \\  
              &              &                &    0.02 &           0.03 &           0.03 &           0.05 &                       &    0.07 &           0.02 &           0.07 &           0.03 &           0.08 &           0.06 &           0.02 &           0.07 &           0.06 &           0.08 &           0.02 &           0.04 \\         
HD 77065 &               &         &  -0.35 &   -0.28 &   -0.16 &   -0.27 &          &  -0.41 &   -0.23 &   -0.33 &   -0.15 &   -0.22 &   -0.21 &   -0.40 &   -0.30 &   -0.53 &   -0.30 &   -0.41 &   -0.21 \\ 
              &              &                &    0.06 &           0.08 &           0.04 &           0.03 &                       &    0.08 &           0.10 &           0.09 &           0.08 &           0.09 &           0.07 &           0.07 &           0.09 &           0.10 &           0.06 &           0.05 &           0.08 \\         
BD+26 1888 &      0.86 &    0.05 &    0.07 &    -0.16 &   0.25 &    0.01 &           &  -0.06 &   0.26 &    -0.02 &   0.12 &    0.09 &    0.20 &    0.03 &    0.05 &    0.13 &    0.10 &    0.04 &    0.09 \\  
              &  0.07 &           0.07 &           0.08 &           0.07 &           0.15 &           0.04 &                       &    0.08 &           0.18 &           0.10 &           0.09 &           0.14 &           0.07 &           0.07 &           0.11 &           0.11 &           0.05 &           0.06 &           0.15 \\         
BD+20 2457 &             &  -0.26 &   -0.81 &   -0.41 &   -0.66 &   -0.50 &          &  -0.65 &   -0.74 &   -0.70 &   -0.49 &   -0.46 &   -0.57 &   -0.75 &   -0.50 &   -1.00 &   -0.65 &   -0.82 &   -0.73 \\ 
              &              &    0.07 &           0.11 &           0.11 &           0.05 &           0.06 &                       &    0.10 &           0.16 &           0.10 &           0.10 &           0.13 &           0.08 &           0.08 &           0.21 &           0.14 &           0.06 &           0.06 &           0.10 \\         
HD 89707 &        -0.62 &          &  -0.34 &   -0.41 &          &  -0.39 &          &  -0.41 &          &  -0.61 &   -0.42 &   -0.50 &   -0.60 &   -0.55 &   -0.40 &   -0.45 &          &  -0.55 &   -0.49 \\ 
              &  0.03 &                       &    0.07 &           0.12 &                       &    0.03 &                       &    0.05 &                       &    0.03 &           0.06 &           0.05 &           0.04 &           0.05 &           0.04 &           0.09 &                       &    0.03 &           0.04 \\         
HD 92320 &        -0.17 &          &  -0.16 &   -0.07 &          &  -0.13 &          &  -0.09 &   -0.13 &   -0.12 &   -0.09 &   -0.06 &   -0.15 &   -0.07 &   -0.02 &   -0.06 &   -0.23 &   -0.14 &   -0.16 \\ 
              &  0.04 &                       &    0.04 &           0.05 &                       &    0.02 &                       &    0.06 &           0.04 &           0.05 &           0.04 &           0.05 &           0.05 &           0.03 &           0.06 &           0.09 &           0.03 &           0.03 &           0.06 \\         
11 Com &          -0.45 &   -0.12 &   -0.05 &   -0.19 &   -0.12 &   -0.16 &   -0.01 &   -0.20 &   -0.43 &   -0.25 &   -0.23 &   -0.35 &   -0.25 &   -0.31 &   -0.24 &   -0.27 &   -0.24 &   -0.30 &   -0.37 \\ 
              &  0.08 &           0.08 &           0.08 &           0.09 &           0.06 &           0.05 &           0.09 &           0.11 &           0.21 &           0.12 &           0.12 &           0.19 &           0.10 &           0.10 &           0.11 &           0.15 &           0.08 &           0.07 &           0.15 \\         
NGC 4349-127 &    -0.20 &   -0.21 &   0.23 &    0.05 &    -0.12 &   0.07 &    0.69 &    -0.14 &          &  -0.23 &   -0.19 &   -0.07 &   -0.06 &   -0.17 &   -0.19 &   0.03 &    -0.13 &   -0.20 &       \\   
              &  0.08 &           0.07 &           0.18 &           0.08 &           0.05 &           0.06 &           0.08 &           0.10 &                       &    0.12 &           0.10 &           0.16 &           0.10 &           0.09 &           0.11 &           0.15 &           0.07 &           0.07 &                       \\  
HD 114762 &       -0.51 &          &  -0.56 &   -0.53 &   -0.45 &   -0.55 &          &  -0.52 &          &  -0.72 &   -0.54 &   -0.54 &   -0.64 &   -0.72 &   -0.57 &   -0.92 &          &  -0.75 &   -0.66 \\ 
              &  0.01 &                       &    0.05 &           0.10 &           0.03 &           0.03 &                       &    0.05 &                       &    0.05 &           0.04 &           0.05 &           0.09 &           0.03 &           0.06 &           0.14 &                       &    0.03 &           0.02 \\         
HD 122562 &       0.40 &    -0.05 &   0.54 &    0.43 &    0.49 &    0.49 &           &  0.22 &           &  0.37 &    0.32 &    0.33 &    0.43 &    0.31 &    0.34 &    0.73 &    0.52 &    0.39 &    0.49 \\  
              &  0.09 &           0.08 &           0.10 &           0.09 &           0.08 &           0.05 &                       &    0.10 &                       &    0.13 &           0.12 &           0.18 &           0.09 &           0.10 &           0.14 &           0.17 &           0.14 &           0.07 &           0.20 \\         
              &  0.07 &                       &    0.13 &           0.07 &           0.05 &           0.04 &                       &    0.09 &           0.13 &           0.09 &           0.09 &           0.10 &           0.07 &           0.07 &           0.09 &           0.12 &           0.10 &           0.06 &           0.09 \\         
HD 132032 &       0.10 &    0.44 &    0.21 &    0.11 &    0.09 &    0.09 &    0.21 &    0.05 &    0.02 &    0.02 &    0.03 &    0.06 &    0.02 &    0.09 &    0.05 &    0.16 &    0.03 &    0.09 &    0.08 \\  
              &  0.02 &           0.03 &           0.08 &           0.02 &           0.03 &           0.02 &           0.03 &           0.04 &           0.03 &           0.04 &           0.02 &           0.03 &           0.04 &           0.02 &           0.07 &           0.03 &           0.07 &           0.02 &           0.11 \\         
HD 131664 &       0.24 &    0.10 &    0.35 &    0.33 &    0.32 &    0.32 &    0.27 &    0.25 &    0.28 &    0.28 &    0.26 &    0.27 &    0.30 &    0.31 &    0.28 &    0.44 &    0.29 &    0.32 &        \\   
              &  0.02 &           0.03 &           0.02 &           0.02 &           0.02 &           0.02 &           0.03 &           0.04 &           0.02 &           0.03 &           0.02 &           0.03 &           0.03 &           0.02 &           0.05 &           0.03 &           0.05 &           0.02 &                       \\  
HD 134113 &       -0.53 &          &  -0.69 &   -0.55 &   -0.56 &   -0.61 &          &  -0.57 &          &  -0.92 &   -0.70 &   -0.68 &   -0.82 &   -0.93 &   -0.83 &   -1.17 &   -0.83 &   -0.90 &   -0.79 \\ 
              &  0.06 &                       &    0.08 &           0.09 &           0.04 &           0.02 &                       &    0.08 &                       &    0.08 &           0.06 &           0.08 &           0.10 &           0.07 &           0.09 &           0.11 &           0.05 &           0.04 &           0.07 \\         
HD 136118 &       -0.58 &          &         &         &         &  -0.20 &   -0.78 &   -0.13 &          &  -0.39 &   -0.01 &   -0.43 &   -0.27 &   -0.10 &          &  -0.11 &   -0.08 &   -0.27 &   -0.81 \\ 
              &  0.04 &                       &                &                &                &    0.09 &           0.04 &           0.06 &                       &    0.06 &           0.09 &           0.10 &           0.11 &           0.09 &                       &    0.07 &           0.10 &           0.05 &           0.05 \\         
HD 137759 &              &  0.27 &    0.41 &    0.36 &    0.40 &    0.37 &           &  0.18 &           &  0.36 &    0.31 &    0.25 &    0.48 &    0.23 &    0.26 &    0.81 &    0.39 &    0.30 &        \\   
              &              &    0.08 &           0.09 &           0.09 &           0.05 &           0.06 &                       &    0.11 &                       &    0.12 &           0.12 &           0.16 &           0.10 &           0.10 &           0.12 &           0.17 &           0.08 &           0.07 &                       \\  
HD 137510 &       0.28 &           &  0.55 &    0.21 &           &  0.35 &    0.20 &    0.24 &           &  0.32 &    0.21 &    0.26 &    0.13 &    0.20 &    0.30 &    0.46 &    0.18 &    0.28 &        \\   
              &  0.03 &                       &    0.03 &           0.09 &                       &    0.02 &           0.03 &           0.06 &                       &    0.05 &           0.03 &           0.05 &           0.07 &           0.02 &           0.03 &           0.05 &           0.07 &           0.02 &                       \\  
HD 140913 &              &  0.56 &    -0.13 &   -0.35 &   -0.16 &   -0.17 &   -0.07 &   -0.55 &          &  -0.23 &   0.02 &    -0.14 &   0.05 &    -0.04 &   0.04 &    0.21 &    0.05 &    -0.11 &   -0.47 \\ 
              &              &    0.04 &           0.06 &           0.04 &           0.04 &           0.10 &           0.04 &           0.18 &                       &    0.05 &           0.08 &           0.11 &           0.08 &           0.14 &           0.10 &           0.06 &           0.15 &           0.06 &           0.05 \\         
HD 156846 &       0.09 &           &  0.24 &    0.15 &           &  0.16 &           &  0.14 &           &  0.12 &    0.10 &    0.12 &    0.07 &    0.11 &    0.17 &    0.16 &    0.08 &    0.13 &    0.04 \\  
              &  0.03 &                       &    0.06 &           0.03 &                       &    0.02 &                       &    0.04 &                       &    0.03 &           0.02 &           0.03 &           0.04 &           0.02 &           0.04 &           0.04 &           0.03 &           0.02 &           0.05 \\         
HD 160508 &       -0.09 &          &  -0.08 &   -0.15 &   -0.30 &   -0.12 &          &  -0.06 &          &  -0.27 &   -0.21 &   -0.21 &   -0.14 &   -0.17 &   -0.13 &   -0.23 &   -0.23 &   -0.19 &   -0.28 \\ 
              &  0.02 &                       &    0.06 &           0.05 &           0.03 &           0.06 &                       &    0.04 &                       &    0.05 &           0.03 &           0.04 &           0.06 &           0.03 &           0.06 &           0.05 &           0.10 &           0.02 &           0.04 \\         
HD 162020 &       0.69 &           &  -0.13 &   -0.19 &   -0.07 &   -0.01 &          &  -0.20 &   0.05 &    -0.13 &   0.03 &    -0.03 &   0.14 &    0.03 &    0.05 &    0.07 &    -0.05 &   -0.06 &   -0.04 \\ 
              &  0.06 &                       &    0.06 &           0.06 &           0.05 &           0.03 &                       &    0.12 &           0.14 &           0.08 &           0.07 &           0.09 &           0.07 &           0.06 &           0.09 &           0.09 &           0.04 &           0.05 &           0.09 \\         
HD 167665 &       -0.09 &          &  -0.16 &   -0.16 &   -0.27 &   -0.16 &   -0.21 &   -0.15 &          &  -0.28 &   -0.24 &   -0.22 &   -0.25 &   -0.24 &   -0.16 &   -0.29 &   -0.35 &   -0.26 &   -0.32 \\ 
              &  0.02 &                       &    0.07 &           0.08 &           0.06 &           0.02 &           0.18 &           0.04 &                       &    0.03 &           0.02 &           0.03 &           0.06 &           0.03 &           0.05 &           0.05 &           0.06 &           0.02 &           0.09 \\         
HD 168443 &       0.14 &    0.12 &    0.04 &    0.20 &    0.19 &    0.10 &    0.01 &    0.04 &    0.02 &    0.08 &    0.08 &    0.15 &    0.03 &    0.02 &    0.03 &    0.04 &    0.04 &    0.01 &    0.17 \\  
              &  0.05 &           0.06 &           0.04 &           0.05 &           0.03 &           0.02 &           0.05 &           0.06 &           0.05 &           0.07 &           0.06 &           0.08 &           0.05 &           0.05 &           0.08 &           0.08 &           0.05 &           0.04 &           0.20 \\         
HD 174457 &       -0.06 &          &  -0.18 &   -0.21 &   -0.09 &   -0.22 &   -0.27 &   -0.18 &          &  -0.29 &   -0.22 &   -0.12 &   -0.16 &   -0.17 &   -0.18 &   -0.28 &   -0.23 &   -0.31 &   -0.42 \\ 
              &  0.02 &                       &    0.06 &           0.06 &           0.08 &           0.01 &           0.11 &           0.04 &                       &    0.06 &           0.02 &           0.13 &           0.05 &           0.06 &           0.08 &           0.07 &           0.02 &           0.02 &           0.06 \\         
HD 175679 &       -0.26 &   -0.09 &   0.19 &    0.09 &    0.01 &    0.07 &    0.20 &    0.02 &    -0.24 &   -0.10 &   -0.08 &   0.01 &    -0.10 &   -0.06 &   0.07 &    0.20 &    -0.09 &   -0.08 &       \\   
              &  0.06 &           0.07 &           0.08 &           0.07 &           0.07 &           0.04 &           0.08 &           0.08 &           0.13 &           0.10 &           0.09 &           0.13 &           0.08 &           0.07 &           0.17 &           0.11 &           0.06 &           0.06 &                       \\  
HD 180314 &       0.09 &    0.13 &           &  0.38 &    0.41 &    0.32 &    0.54 &    0.19 &    0.29 &    0.31 &    0.24 &    0.31 &    0.27 &    0.20 &    0.30 &    0.79 &    0.35 &    0.26 &    0.36 \\  
              &  0.09 &           0.09 &                       &    0.09 &           0.13 &           0.06 &           0.13 &           0.12 &           0.20 &           0.13 &           0.13 &           0.17 &           0.10 &           0.10 &           0.13 &           0.16 &           0.08 &           0.08 &           0.19 \\         
KOI-415 &        0.24 &           &  0.17 &           &  0.61 &    0.06 &           &  0.16 &    0.30 &    0.24 &    0.36 &           &  0.60 &    0.19 &           &  -0.31 &   0.48 &    0.08 &    0.34 \\  
              &  0.19 &                       &    0.15 &                       &    0.10 &           0.06 &                       &    0.09 &           0.05 &           0.07 &           0.07 &                       &    0.08 &           0.07 &                       &    0.10 &           0.13 &           0.05 &           0.06 \\         
HD 190228 &       -0.38 &          &  -0.24 &   -0.19 &   -0.10 &   -0.26 &   -0.12 &   -0.28 &   -0.38 &   -0.31 &   -0.27 &   -0.17 &   -0.31 &   -0.36 &   -0.38 &   -0.25 &   -0.29 &   -0.36 &   -0.12 \\ 
              &  0.07 &                       &    0.10 &           0.06 &           0.11 &           0.04 &           0.07 &           0.08 &           0.06 &           0.12 &           0.09 &           0.17 &           0.08 &           0.07 &           0.09 &           0.12 &           0.05 &           0.06 &           0.10 \\         
HR 7672 &         0.11 &           &  -0.02 &   -0.02 &   0.10 &    0.02 &    0.00 &    0.00 &    -0.02 &   -0.10 &   -0.05 &   -0.07 &   0.01 &    0.00 &    0.07 &    0.02 &    -0.13 &   -0.04 &   -0.02 \\ 
              &  0.04 &                       &    0.04 &           0.05 &           0.16 &           0.02 &           0.14 &           0.05 &           0.04 &           0.05 &           0.04 &           0.05 &           0.05 &           0.04 &           0.09 &           0.06 &           0.05 &           0.03 &           0.11 \\         
HD 191760 &       0.24 &    0.44 &    0.35 &    0.30 &    0.28 &    0.28 &    0.18 &    0.23 &    0.22 &    0.27 &    0.20 &    0.23 &    0.22 &    0.25 &    0.21 &    0.36 &    0.23 &    0.27 &    0.22 \\  
              &  0.01 &           0.03 &           0.01 &           0.01 &           0.02 &           0.01 &           0.05 &           0.04 &           0.01 &           0.02 &           0.01 &           0.03 &           0.02 &           0.01 &           0.04 &           0.03 &           0.04 &           0.01 &           0.08 \\         
HD 202206 &       0.16 &    0.14 &    0.30 &    0.29 &    0.28 &    0.29 &    0.18 &    0.22 &    0.23 &    0.25 &    0.24 &    0.27 &    0.29 &    0.30 &    0.29 &    0.43 &    0.25 &    0.29 &    0.17 \\  
              &  0.04 &           0.04 &           0.02 &           0.03 &           0.02 &           0.02 &           0.07 &           0.04 &           0.03 &           0.04 &           0.03 &           0.04 &           0.04 &           0.03 &           0.04 &           0.05 &           0.04 &           0.02 &           0.12 \\         
HD 209262 &       0.04 &    -0.17 &   0.08 &    0.17 &    0.12 &    0.07 &    0.14 &    0.04 &    0.00 &    0.04 &    0.03 &    0.06 &    0.04 &    0.07 &    0.08 &    0.13 &    0.04 &    0.05 &    0.02 \\  
              &  0.05 &           0.04 &           0.02 &           0.06 &           0.02 &           0.02 &           0.04 &           0.04 &           0.04 &           0.04 &           0.03 &           0.04 &           0.03 &           0.03 &           0.04 &           0.04 &           0.04 &           0.02 &           0.10 \\         
BD+24 4697 &      0.42 &           &  -0.38 &   -0.26 &   -0.16 &   -0.18 &          &  -0.23 &   -0.14 &   -0.30 &   -0.10 &   -0.15 &   -0.07 &   -0.17 &   -0.10 &   -0.19 &   -0.06 &   -0.20 &   -0.27 \\
HD 217786 &       -0.08 &   -0.25 &   -0.12 &   -0.11 &   -0.16 &   -0.15 &          &  -0.15 &          &  -0.25 &   -0.22 &   -0.20 &   -0.23 &   -0.19 &   -0.21 &   -0.27 &   -0.24 &   -0.21 &   -0.23 \\ 
              &  0.01 &           0.03 &           0.02 &           0.05 &           0.03 &           0.01 &                       &    0.04 &                       &    0.03 &           0.01 &           0.02 &           0.03 &           0.01 &           0.08 &           0.03 &           0.04 &           0.01 &           0.05 \\         
HD 219077 &       -0.13 &   -0.06 &   -0.21 &   0.00 &    -0.03 &   -0.10 &   0.03 &    -0.13 &   -0.22 &   -0.14 &   -0.11 &   -0.07 &   -0.19 &   -0.18 &   -0.12 &   -0.19 &   -0.20 &   -0.22 &       \\   
              &  0.06 &           0.07 &           0.06 &           0.06 &           0.05 &           0.03 &           0.07 &           0.08 &           0.07 &           0.09 &           0.08 &           0.09 &           0.07 &           0.07 &           0.09 &           0.10 &           0.06 &           0.05 &                       \\ 
\end{longtable}
\end{landscape}
\end{longtab}
}

\section{Results}
\label{results_sect}
\subsection{Abundance patterns of stars with brown dwarfs}
\label{comparison_swbds_mass}

  Recent studies \citep{2011A&A...525A..95S,2014MNRAS.439.2781M,2014A&A...566A..83M}
  have suggested that the formation mechanisms of BDs
  companions with masses above and below  $\sim$ 42.5 M$_{\rm Jup}$
  may be different. Therefore, we have divided the sample of stars with brown dwarfs
  in  stars with BDs candidates in the mass range M$_{\rm C}$$\sin i$ $<$ 42.5 M$_{\rm Jup}$
  and stars with BDs candidates with masses M$_{\rm C}$$\sin i$ $>$ 42.5 M$_{\rm Jup}$
  \footnote{ Note that through the text we use the notation ``minimum mass
  M$_{\rm C}$$\sin i$'', however there are seven candidates with true mass determinations 
  (four in the mass range M$_{\rm C}$ $>$ 42.5 M$_{\rm Jup}$, and three with
   M$_{\rm C}$ $<$ 42.5 M$_{\rm Jup}$).}.

  The SWBDs with candidates in the mass range M$_{\rm C}$$\sin i$ $<$ 42.5 M$_{\rm Jup}$ is composed
  of 32 stars, including 4 F stars, 19 G stars and 9 K stars. Regarding its evolutionary state 8 are 
  giants, 12 are classified as subgiants, and 12 stars are on the main-sequence.
  The total number of stars in the SWBDs with M$_{\rm C}$$\sin i$ $>$ 42.5 M$_{\rm Jup}$ is 21,
  including 6 F stars, 12 G stars and 3 K stars. The sample is composed of 14 main-sequence stars
  as well as 7 stars in the subgiant branch. 
  In the following we compare the metallicities and abundances 
  of these two subsamples. 

\subsubsection{Stellar biases}
\label{possible_biases}

\begin{figure}[htb]
\centering
\includegraphics[angle=270,scale=0.45]{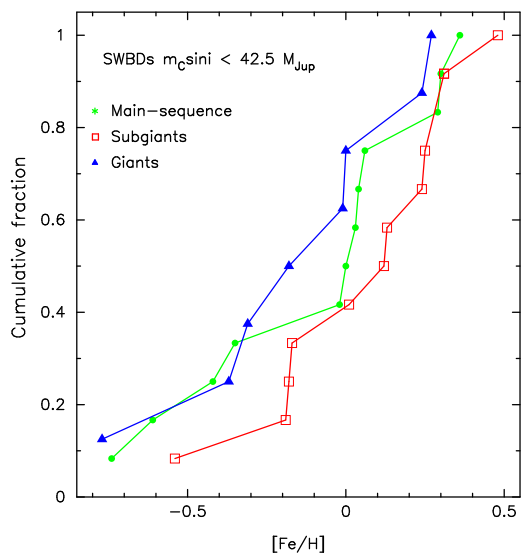}
\caption{
Histogram of cumulative frequencies for SWBDs with companions in the mass range
M$_{\rm C}$$\sin i$ $<$ 42.5 M$_{\rm Jup}$
according to their luminosity class.
}
\label{dist_acum_low_mass_lc}
\end{figure}

   Before a comparison of the metallicities and individual
   abundances between the two defined samples of stars with brown dwarfs candidates 
  is done,
  a comparison of the stellar properties
  of both samples was performed, in particular in terms of age, distance, and kinematics,
  since these parameters are most likely to reflect the original metal content of
  the molecular cloud where the stars were born. The comparison is shown
  in Table~\ref{bias_table}. A series of two-sided Kolmogorov-Smirnov (K-S)
  tests \citep[e.g.][]{1983MNRAS.202..615P} were performed to check whether the samples
  are likely or not drawn from the same parent population. 
  The comparison shows that both samples show similar distributions in brightness and age.
  The sample of SWBDs with M$_{\rm C}$$\sin i$ $<$ 42.5 M$_{\rm Jup}$ contains stars 
  out to larger distances and slightly larger stellar masses than the SWBDs with M$_{\rm C}$$\sin i$ $>$ 42.5 M$_{\rm Jup}$.
  Nevertheless, we note that the median distance for both samples are quite similar
  and most of the stars,  $\sim$ 75\% with M$_{\rm C}$$\sin i$ $<$ 42.5 M$_{\rm Jup}$ are within 92 pc
  (the volume covered by the SWBDs with companions above 42.5 M$_{\rm Jup}$).
  As a further check, the metallicity-distance plane was explored finding
  no metallicity difference between the SWBDs with M$_{\rm C}$$\sin i$ $<$ 42.5 M$_{\rm Jup}$ 
  located closer and farther than 92 pc. 
   This potential bias is discussed with more detail in Sec.~\ref{distance_bias}.
  Regarding the stellar mass, only four SWBDs with M$_{\rm C}$$\sin i$ $<$ 42.5 M$_{\rm Jup}$
  show stellar masses larger than 1.4 M$_{\odot}$
  (SWBDs with companions above 42.5 M$_{\rm Jup}$ cover up to 1.31 M$_{\odot}$)
  showing a large range of metallicities
  (two stars have [Fe/H] $\sim$ +0.25, one shows solar metallicity, and the other one
  is metal-poor with [Fe/H] $\sim$ -0.30).  
  We note that in the sample of SWBDs with M$_{\rm C}$$\sin i$ $>$ 42.5 M$_{\rm Jup}$
  there are no giant stars, indeed nearly all stars ($\sim$ 67\%) are in the main-sequence.
  However, in its less massive counterpart sample, most of the stars are evolved,
  with about 62.5\% of the stars in the giant and subgiant phase.
   This fact should be analysed carefully, since it has been shown  that unlike their
  main-sequence counterparts, it is still unclear
  whether giant stars with planets show or not metal-enrichment
  \citep{
  2005PASJ...57..127S,2005ApJ...632L.131S,2007A&A...475.1003H,2007A&A...473..979P,2008PASJ...60..781T,2010ApJ...725..721G,
  2013A&A...554A..84M,2013A&A...557A..70M,2015A&A...574A..50J,2015A&A...574A.116R,2016A&A...588A..98M}.
  Further, the abundance of some elements might be influenced
  by 3D or nLTE effects \citep[e.g.][]{2011JPhCS.328a2002B,2011A&A...528A..87M}.
  The metallicity distribution of SWBDs with M$_{\rm C}$$\sin i$ $<$ 42.5 M$_{\rm Jup}$
  is shown in Figure~\ref{dist_acum_low_mass_lc} where the stars are classified according to their luminosity
  class. The figure does not reveal any clear difference in metallicity between
  giants, subgiants, and main-sequence stars. 
  We will discuss this issue in more detail in Sec.~\ref{giants_planets}.

\subsubsection{Kinematic biases}
\label{kinematic_biases}

  Regarding kinematics, stars were classified as belonging to the
  thin/thick disc applying the methodology described in 
  \citet{2003A&A...410..527B,2005A&A...433..185B}. For this purpose,
  we first computed the stellar spatial Galactic velocity components
  $(U,V,W)$ using the methodology described in
  \citet{2001MNRAS.328...45M} and \citet{2010A&A...521A..12M},
  using the {\sc Hipparcos} parallaxes \citep{Leeuwen} and Tycho-2 proper motions
  \citep{2000A&A...355L..27H}. Radial velocities were taken from the compilation of
  \cite{2007AN....328..889K}. 
  The results show that most of the stars in both subsamples
  ($\sim$ 72\%, and $\sim$ 75\%, respectively) should, according to their
  kinematics, belong to the thin disc\footnote{We note that
  our objective here is to discard the presence of a significant
  fraction of thick-disc stars within our samples
  (as these stars are expected to be relatively old, metal poor,
  and to show $\alpha$-enhancement) and
  not a detailed thin/thick disc classification which
  would require a detailed analysis of kinematics,
  ages, and abundances.}. 

  Another potential bias comes from the fact that several stars might harbour
  additional companions in the planetary range. 
  Five SWBDs in the mass domain M$_{\rm C}$$\sin i$ $<$ 42.5 M$_{\rm Jup}$
  are known to host, in addition to a brown dwarf,  at least one companion in
  the gas-giant planetary mass domain.
  These stars are HD 38529, HD 168443, HIP 5158, and HAT-P-13
  (all with the planet closer to the star than the brown dwarf),
  and HD 2022206,
 where the brown dwarf occupies the innermost orbit.
  We note that all these stars, except
  one (HD 168443), show significant positive metallicities.
  In order to test whether this fact could affect our results 
  we compared the metallicity distribution of SWBDs
  with M$_{\rm C}$$\sin i$ $<$ 42.5 M$_{\rm Jup}$ when all the 32 stars
  with companions in this mass range are considered and when the four stars with possible additional planets
  are excluded. 
  The results from the K-S test show that both distributions are
  virtually equal with a probability of $\sim$ 99\%.
  Further analysis of this potential bias will be 
  provided in Sec.~\ref{giants_planets}.
  Finally, we note that only one star (BD+20 2457) harbours
  two companions in the brown dwarf regime.

\begin{table*}[!htb]
\centering
\caption{
Comparison between the properties of SWBDs with M$_{\rm C}$$\sin i$ $<$ 42.5 M$_{\rm Jup}$
and SWBDs with M$_{\rm C}$$\sin i$ $>$ 42.5 M$_{\rm Jup}$.
}
\label{bias_table}
\begin{tabular}{lllllllll}
\hline\noalign{\smallskip}
               &  \multicolumn{3}{c}{\textbf{M$_{\rm C}$$\sin i$ $<$ 42.5 M$_{\rm Jup}$}}  & \multicolumn{3}{c}{\textbf{M$_{\rm C}$$\sin i$ $>$ 42.5 M$_{\rm Jup}$}} & \multicolumn{2}{c}{\textbf{K-S test}} \\
               &  \multicolumn{3}{c}{\hrulefill}      & \multicolumn{3}{c}{\hrulefill}     & \multicolumn{2}{c}{\hrulefill}  \\
               & Range       & Mean        &  Median  &   Range       & Mean     & Median  &  $D$    & $p$  \\
\hline\noalign{\smallskip}
V (mag)            & 3.29/10.82 & 7.70   &  7.78  &     5.80/9.77   & 7.70  & 7.68   &     0.24 &   0.40 \\
 Distance (pc) & 18.3/2174  & 166.9  & 46.58  &     17.8/92.3   & 44.4  & 44.9   &     0.26 &   0.31 \\
Age (Gyr)          & 0.66/11.48 & 5.07   & 4.33   &     0.78/11.13  & 5.31  & 5.29   &     0.25 &   0.43 \\
Mass (M$_{\odot})$ & 0.40/2.53  &  1.15  &  1.10  &     0.62/1.31   &  0.97 &  0.99  &     0.31 &   0.16 \\ 
T$_{\rm eff}$ (K)  & 4168/6163  & 5330   & 5570   &     4860/6240   &  5697 &  5795  &     0.34 &   0.08  \\
SpType(\%)         & \multicolumn{3}{l}{13 (F); 59 (G); 28 (K) } & \multicolumn{3}{l}{29 (F); 57 (G); 14 (K)} & \multicolumn{2}{c}{  } \\
LC(\%)$^{\dag}$    & \multicolumn{3}{l}{25 (G); 37.5 (S); 37.5 (MS)} & \multicolumn{3}{l}{33 (S); 67 (MS)      } & \multicolumn{2}{c}{  } \\
D/TD(\%)$^{\ddag}$ & \multicolumn{3}{l}{72 (D); 9 (TD); 19 (R)} & \multicolumn{3}{l}{75 (D); 10 (TD); 15 (R)} & \multicolumn{2}{c}{  } \\
\noalign{\smallskip}\hline\noalign{\smallskip} 
\multicolumn{9}{l}{$^{\dag}$ MS: Main-sequence, S: Subgiant, G: Giant} \\
\multicolumn{9}{l}{$^{\ddag}$ D: Thin disc, TD: Thick disc, R: Transition} \\
\end{tabular}
\end{table*}

\subsubsection{Metallicity distributions}
\label{metallicity_distributions}

\begin{figure}[htb]
\centering
\includegraphics[angle=270,scale=0.55]{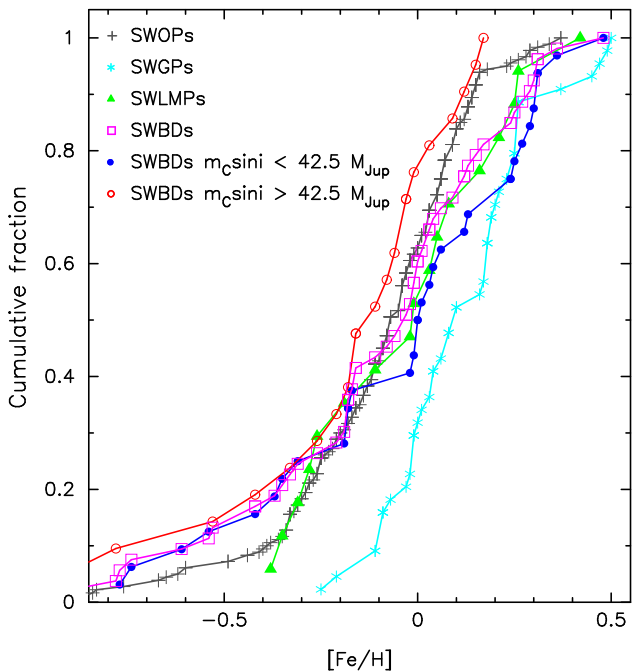}
\caption{
Histogram of cumulative frequencies for the different samples
studied in this work.
}
\label{distribuciones_acumuladas}
\end{figure}

 As mentioned before  32 SWBDs 
 are in the mass range M$_{\rm C}$$\sin i$ $<$ 42.5 M$_{\rm Jup}$ whilst
 21 stars host BDs candidates with masses M$_{\rm C}$$\sin i$ $>$ 42.5 M$_{\rm Jup}$.
 Some statistical diagnostics for both samples are summarised in
 Table~\ref{metal_statistics}, while their metallicity cumulative distribution functions
 are shown in Figure~\ref{distribuciones_acumuladas}. 
 We also show the metallicity distribution of the whole sample of stars with
 brown dwarfs (i.e., all the 53 stars with brown dwarf companions, SWBDs).
 In addition, several samples are overplotted for comparison: {\it i)} a sample of stars
 without known planetary companions (180 stars, SWOPs), {\it ii)} a sample
 of stars with known gas-giant planets (44 stars, SWGPs), and
 {\it iii)} a sample of stars with known low-mass planets ( M$_{\rm p}\sin i$ $<$ 30 M$_{\oplus}$, 17 stars,
 SWLMPs). In order to be as homogeneous as possible,
 these comparison samples were taken from 
 \cite{2015A&A...579A..20M} so their stellar parameters are 
 determined with the same technique used in this work and using
 similar spectra.

\begin{table}
\centering
\caption{[Fe/H] statistics of the stellar samples.}
\label{metal_statistics}
\begin{scriptsize}
\begin{tabular}{lcccccc}
\hline\noalign{\smallskip}
 Sample  &  $Mean$ & $ Median$ &  $\sigma$  &  $Min$&  $Max$ & $N$ \\
\hline\noalign{\smallskip}
SWBDs                  & -0.10 & -0.03 & 0.32 & -0.92 & 0.48 & 53 \\ 
BDs M$_{\rm C}$$\sin i$ $<$ 42.5 M$_{\rm Jup}$ & -0.04 &  0.01 & 0.33 & -0.77 & 0.48 & 32 \\
BDs M$_{\rm C}$$\sin i$ $>$ 42.5 M$_{\rm Jup}$ & -0.18 & -0.11 & 0.28 & -0.92 & 0.17 & 21  \\
\hline
SWOPs                  & -0.10 & -0.07 & 0.24 & -0.87 & 0.37 & 180 \\
SWGPs                  &  0.12 &  0.10 & 0.18 & -0.25 & 0.50 &  44 \\
SWLMPs                 & -0.03 & -0.01 & 0.23 & -0.38 & 0.42 &  17 \\
\noalign{\smallskip}\hline\noalign{\smallskip}
\end{tabular}
\end{scriptsize}
\end{table}

  There are a few interesting facts to be taken from the
 distributions shown in Figure~\ref{distribuciones_acumuladas}: 
 {\it i)} SWBDs as a whole  (magenta line) do not follow the well trend of SWGPs
 (light-blue line) of showing metal-enrichment;
 {\it ii)} considering the global metallicity distribution of SWBDs,
   there is a trend of SWBDs in the mass domain M$_{\rm C}$$\sin i$ $<$ 42.5 M$_{\rm Jup}$
 (dark-blue line)
 of having larger metallicities than SWBDs with M$_{\rm C}$$\sin i$ $>$ 42.5 M$_{\rm Jup}$
 (red line); 
 {\it iii)} for metallicities below approximately -0.20 the metallicity distributions 
 of SWBDs with masses above and below 42.5 M$_{\rm Jup}$ 
 seem to follow a similar trend; 
 {\it iv)} for larger metallicities the
 distribution of SWBDs with companions in the mass range M$_{\rm C}$$\sin i$ $<$ 42.5 M$_{\rm Jup}$
 clearly shifts towards higher metallicities
  when compared with the distribution of SWBDs in the mass range
 M$_{\rm C}$$\sin i$ $>$ 42.5 M$_{\rm Jup}$.
 We also note that at high-metallicities,
 (larger than +0.20), the metallicity distribution of
 of SWBDs in the mass domain M$_{\rm C}$$\sin i$ $<$ 42.5 M$_{\rm Jup}$
 is similar to that of SWGPs.


 These results can be compared with previous studies.  
 \cite{2014MNRAS.439.2781M} and \cite{2014A&A...566A..83M}  found that the stars with brown dwarfs
 companions do not show the metal-rich signature seen in 
 stars hosting gas-giant planets. Further,
 \cite{2014MNRAS.439.2781M}  did not report metallicity differences between
 stars with BDs with minimum masses lower and larger
 than 42.5 M$_{\rm Jup}$. 
  We also note that Figure~6 in \cite{2014MNRAS.439.2781M} shows results similar to ours:
 Around metallicities $\sim$ +0.00 stars with BDs with  M$_{\rm C}$$\sin i$ $<$ 42.5 M$_{\rm Jup}$
 tend to show larger metallicities ``reaching'' the metallicity distribution
 of stars with gas-giant planets at [Fe/H] $\sim$ +0.25.

\subsubsection{Other chemical signatures}

\begin{figure}[!htb]
\centering
\includegraphics[angle=270,scale=0.45]{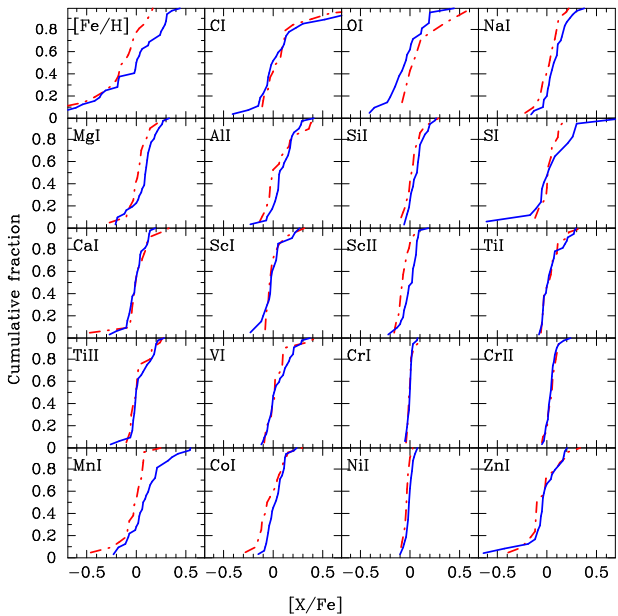}
\caption{
[X/Fe] cumulative fraction of SWBDs with M$_{\rm C}$$\sin i$ $<$ 42.5 M$_{\rm Jup}$
(blue continuous line)  and M$_{\rm C}$$\sin i$ $>$ 42.5 M$_{\rm Jup}$ 
(red dash-to-dot line).
}
\label{distribuciones_acumuladas_abundancias}
\end{figure}

 In order to try to disclose differences in the abundances of
 other elements besides iron, Figure~\ref{distribuciones_acumuladas_abundancias}
 compares the cumulative distribution of [X/Fe]
 between SWBDs with M$_{\rm C}$$\sin i$ below and above 
 42.5 M$_{\rm Jup}$.
 Table~\ref{abundance_statistics} gives some statistic diagnostics,
 the results of a K-S test for each ion
  and also for [X$_{\alpha}$/Fe], [X$_{\rm Fe}$/Fe], and [X$_{\rm vol}$/Fe]
 (see definitions below).
 For Ca~{\sc i}, Sc~{\sc i}, Ti~{\sc i}, Cr~{\sc i}, and Cr~{\sc ii}
 the distributions of both samples seem to be quite similar.
 Indeed the the probabilities of both samples coming from the same 
 distribution returned by the K-S tests for 
 these ions 
 are high ($>$ 80\%).
 On the other hand, for the abundances of Sc~{\sc ii}, Mn~{\sc i},  Ni~{\sc i},
  and X$_{\rm Fe}$ the tests
 conclude that both samples might be different.


\begin{table}
\centering
\caption{Comparison between the elemental abundances of stars with BDs with minimum
masses M$_{\rm C}$$\sin i$ $<$ 42.5 M$_{\rm Jup}$ and M$_{\rm C}$$\sin i$ $>$ 42.5 M$_{\rm Jup}$.
}
\label{abundance_statistics}
\begin{scriptsize}
\begin{tabular}{lccccccc}
\hline\noalign{\smallskip}
 [X/Fe] & \multicolumn{2}{c}{M$_{\rm C}$$\sin i$ $<$ 42.5 M$_{\rm Jup}$} &  \multicolumn{2}{c}{M$_{\rm C}$$\sin i$ $>$ 42.5 M$_{\rm Jup}$} & \multicolumn{3}{c}{K-S test} \\
     &  \multicolumn{2}{c}{\hrulefill} &  \multicolumn{2}{c}{\hrulefill} &  \multicolumn{3}{c}{\hrulefill} \\
     &   Median  &  $\sigma$ & Median &  $\sigma$ & $D$ & $p$-value & n$_{\rm eff}$ \\
\hline\noalign{\smallskip}
  C~{\sc i}     &    0.00     &    0.29    &         0.07   &      0.32   &      0.19  &       0.80    &    11.15  \\
  O~{\sc i}     &   -0.04     &    0.21    &         0.06   &      0.26   &      0.38  &       0.34    &     5.25  \\
 Na~{\sc i}     &    0.10     &    0.12    &         0.03   &      0.11   &      0.26  &       0.32    &    12.35  \\
 Mg~{\sc i}     &    0.10     &    0.14    &         0.03   &      0.13   &      0.40  &       0.03    &    12.16  \\
 Al~{\sc i}     &    0.07     &    0.13    &         0.00   &      0.17   &      0.35  &       0.09    &    11.48  \\
 Si~{\sc i}     &    0.08     &    0.08    &         0.03   &      0.10   &      0.29  &       0.19    &    12.68  \\
  S~{\sc i}     &    0.02     &    0.30    &         0.01   &      0.09   &      0.29  &       0.53    &     6.68  \\
 Ca~{\sc i}     &    0.01     &    0.10    &         0.01   &      0.15   &      0.13  &       0.97    &    12.68  \\
 Sc~{\sc i}     &   -0.02     &    0.12    &        -0.01   &      0.12   &      0.20  &       0.87    &     7.88  \\
Sc~{\sc ii}     &    0.03     &    0.09    &        -0.07   &      0.07   &      0.45  &       0.01    &    12.68  \\
 Ti~{\sc i}     &    0.02     &    0.11    &         0.04   &      0.11   &      0.12  &       0.99    &    12.68  \\
Ti~{\sc ii}     &    0.00     &    0.12    &         0.01   &      0.10   &      0.21  &       0.62    &    12.31  \\
  V~{\sc i}     &    0.02     &    0.13    &         0.02   &      0.13   &      0.20  &       0.63    &    12.68  \\
 Cr~{\sc i}     &    0.00     &    0.03    &         0.00   &      0.04   &      0.17  &       0.85    &    12.68  \\
Cr~{\sc ii}     &    0.04     &    0.06    &         0.06   &      0.05   &      0.15  &       0.94    &    12.16  \\
 Mn~{\sc i}     &    0.09     &    0.19    &         0.00   &      0.15   &      0.45  &       0.01    &    12.68  \\
 Co~{\sc i}     &    0.03     &    0.09    &        -0.03   &      0.14   &      0.37  &       0.05    &    12.16  \\
 Ni~{\sc i}     &    0.00     &    0.04    &        -0.03   &      0.03   &      0.48  &     $<$0.01   &    12.68  \\
 Zn~{\sc i}     &   -0.04     &    0.19    &        -0.10   &      0.17   &      0.36  &       0.09    &    11.20  \\
\hline
$X_{\alpha}$  &  0.04 &        0.09 &         0.01  &       0.10   &      0.23   &      0.45  &     12.68 \\
$X_{\rm Fe}$  &  0.03 &        0.06 &        -0.03  &       0.04   &      0.47   &   $<$0.01  &     12.68 \\
$X_{\rm vol}$ &  0.05 &        0.15 &         0.03  &       0.12   &      0.19   &      0.73  &     12.68 \\
\noalign{\smallskip}\hline\noalign{\smallskip}
\end{tabular}
\end{scriptsize}
\end{table}

 In order to compare with the SWOP, SWGP, and SWLMP samples defined in 
 Sec.~\ref{metallicity_distributions}, we grouped the 
 ions into three categories: alpha elements, iron-peak elements,
 and volatile elements. 
 For alpha and iron-peak elements we follow \cite{2014A&A...566A..83M}
 and define  [X$_{\alpha}$/Fe] as the mean of the [X/Fe] abundances
 of Mg~{\sc i}, Si~{\sc i}, Ca~{\sc i}, and Ti~{\sc i}, while
 [X$_{\rm Fe}$/Fe] is defined as the mean of the
 Cr~{\sc i}, Mn~{\sc i}, Co~{\sc i}, and Ni~{\sc i} abundances.
 We define the mean volatile abundance, [X$_{\rm vol}$/Fe] as
 the mean of the [X/Fe] values of the elements with a condensation
 temperature, T$_{\rm C}$, lower than 900 K namely, C~{\sc i}, O~{\sc i}, S~{\sc i}, and Zn~{\sc i}.
 Although  Na~{\sc i} has a T$_{\rm C}$ slightly above 900 K we include it in the group of volatiles 
 to account for the fact that 
 for some stars the abundances of some volatiles were not obtained. 
 It is important to mention at this point that the abundances in the 
 comparison samples  for this work were derived in a similar way 
 by \cite{2015A&A...579A..20M,2016A&A...588A..98M}.
 
\begin{figure*}[htb]
\centering
\begin{minipage}{0.32\linewidth}
\includegraphics[angle=270,scale=0.38]{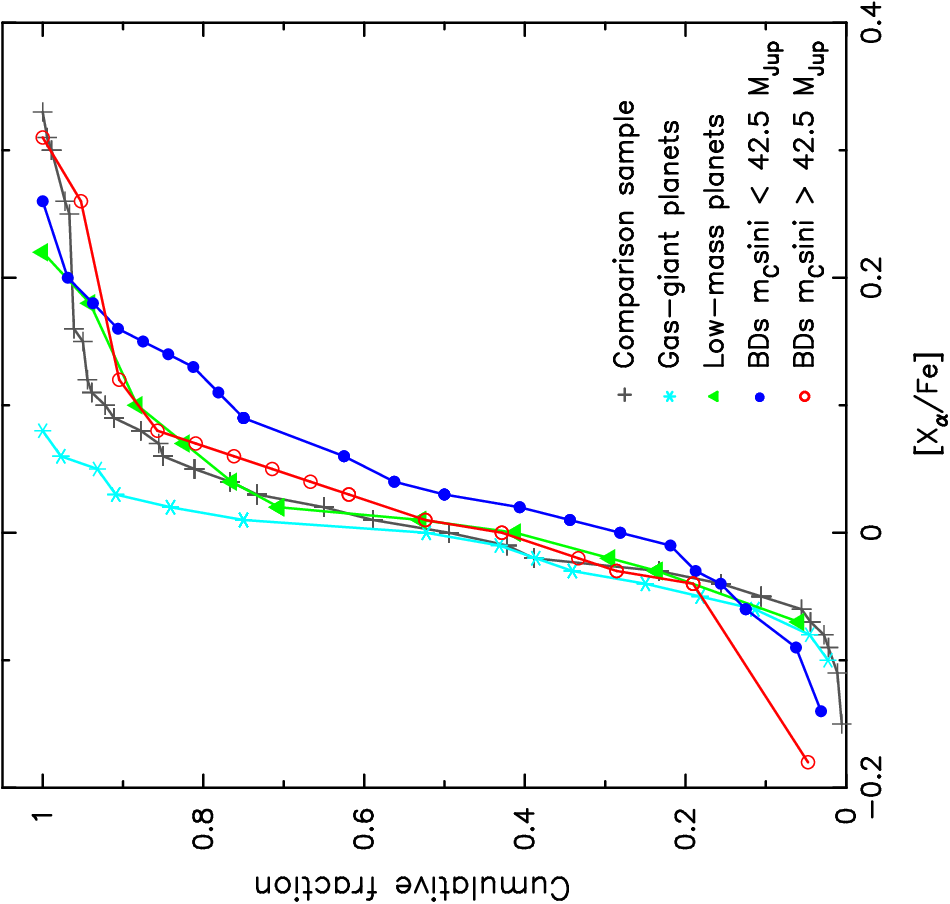}
\end{minipage}
\begin{minipage}{0.32\linewidth}
\includegraphics[angle=270,scale=0.38]{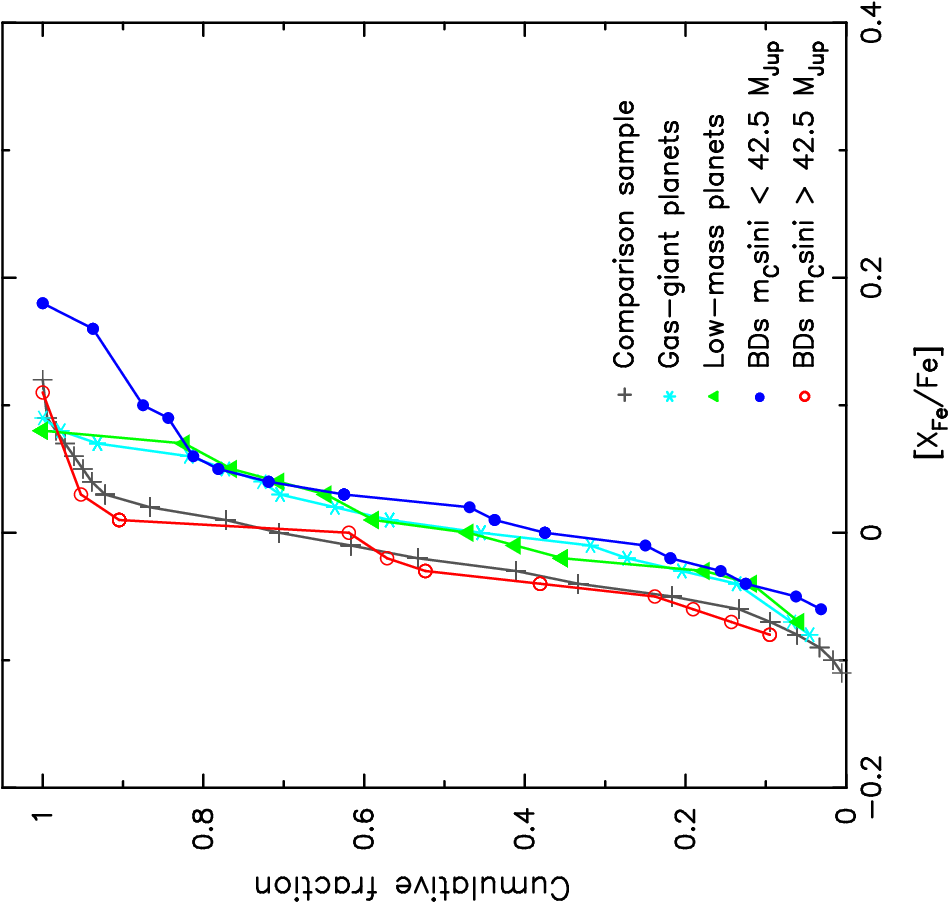}
\end{minipage}
\begin{minipage}{0.32\linewidth}
\includegraphics[angle=270,scale=0.38]{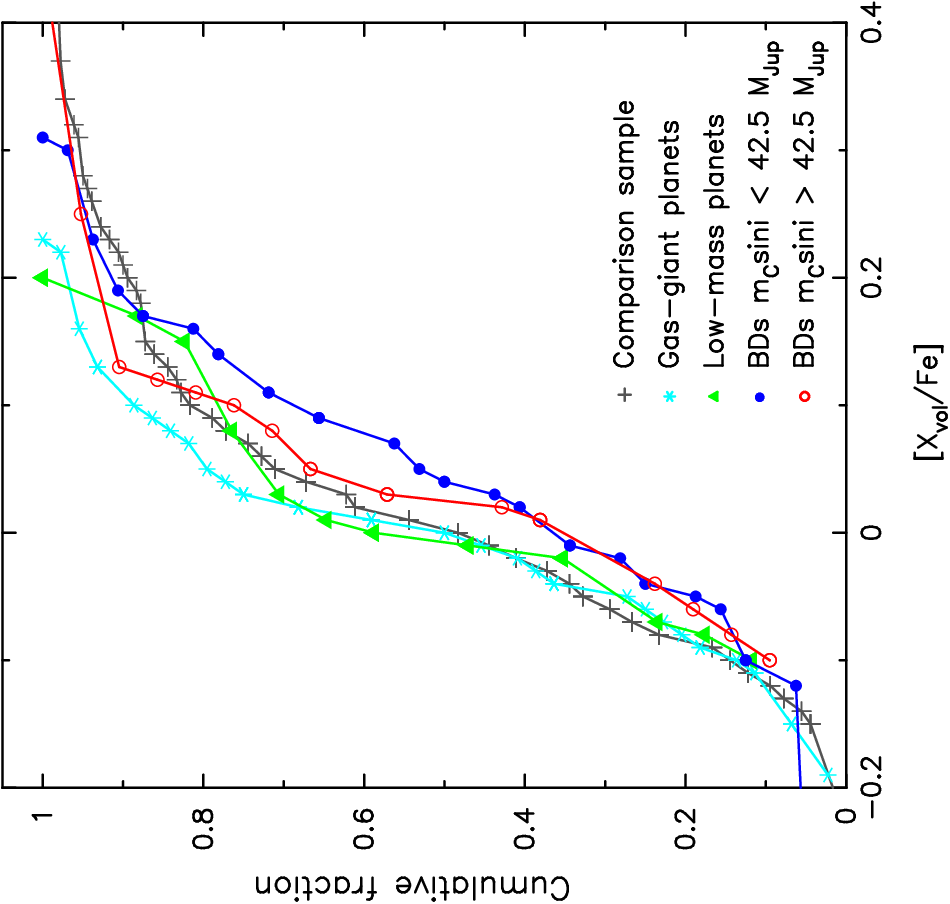}
\end{minipage}
\caption{
 Histogram of cumulative frequencies of [X$_{\alpha}$/Fe] (left), 
 [X$_{\rm Fe}$/Fe] (middle), and [X$_{\rm vol}$/Fe] (right) for the different samples studied in this work.
}
\label{distrib_acu_alpha_iron_vol}
\end{figure*}

 The different cumulative functions are shown in Figure~\ref{distrib_acu_alpha_iron_vol}.
 Interestingly, the figure reveals a tendency of SWBDs
 in the low-mass domain to have slightly larger abundances than
 the rest of the samples in all categories. In order to test this tendency the
 SWBDs with companions in the mass range M$_{\rm C}$$\sin i$ $<$ 42.5 M$_{\rm Jup}$
 was compared (by means of a K-S test) with the rest of the samples.
 The results are provided in Table~\ref{abundance_statistics2}.

 Regarding $\alpha$ elements, the sample of stars with low-mass BDs companions
 does not seem to be different from the SWOP and SWLMP samples, although we note the low $p$-value
 of 0.05 when comparing with the SWOP sample.
 The K-S test suggests, however, that the sample differs from the one
 of stars harbouring gas-giant planets ($p$-value $<$ 0.01). Since this is somehow a surprising result, 
 we have checked if our SWGP sample occupies the same place in the
 [X$_{\alpha}$/Fe] vs. [Fe/H] plot as other samples in the literature.
  For this check we have taken the data from \cite{2012A&A...545A..32A} finding consistent results,
 i.e, both our SWGP and the stars with giant planets from
 \cite{2012A&A...545A..32A} tend to show high metallicity values and rather low [X$_{\alpha}$/Fe] values.
 The most significant differences appear when considering the iron-peak
 elements. In this case the sample of SWBDs in the low-mass companion range 
 seems to be shifted towards higher metallicities when compared with the
 SWOPs. 
 No statistically significant differences are found when
 considering the volatile elements, although we note that in the comparison with 
 the SWGPs the $p$-value is relatively low
 (of only 0.04). 

 We therefore conclude that the SWBD sample with companions in the mass
 range M$_{\rm C}$$\sin i$ $<$ 42.5 M$_{\rm Jup}$ may differ from the
 SWOPs in iron-peak elements, but also from the GWPs when considering
 $\alpha$ elements. 
 
\begin{table}
\centering
\caption{  Comparison between the abundances of stars with BDs with minimum
masses M$_{\rm C}$$\sin i$ $<$ 42.5 M$_{\rm Jup}$ and the
stars without planets, stars with low-mass planets, and stars with gas-giant planets
samples.}
\label{abundance_statistics2}
\begin{tabular}{lcccccc}
\hline\noalign{\smallskip}
         &  \multicolumn{2}{c}{Stars without} & \multicolumn{2}{c}{Stars with low-} & \multicolumn{2}{c}{Stars with gas-} \\
         &  \multicolumn{2}{c}{planets}       & \multicolumn{2}{c}{mass planets}    & \multicolumn{2}{c}{giant planets} \\
$[X/Fe]$ &   $D$ & $p$-value          & $D$ & $p$-value            &  $D$ & $p$-value          \\
\hline
$X_{\alpha}$  & 0.25   &      0.05 & 0.30  &      0.23 &  0.43  & $<$   0.01 \\
$X_{\rm Fe}$  & 0.40   & $<$  0.01 & 0.19  &      0.78 &  0.17  &       0.63 \\
$X_{\rm vol}$ & 0.21   &      0.16 & 0.30  &      0.21 &  0.31  &       0.04 \\
\noalign{\smallskip}\hline\noalign{\smallskip}
\end{tabular}
\tablefoot{n$_{\rm eff}$ $\sim$ 27.2 (SWOPs); $\sim$ 11.1 (SWLMPs); $\sim$ 18.5 (SWGPs).}
\end{table}

\subsubsection{Presence of red giants and additional planetary companions}
\label{giants_planets}
 
 As already pointed out, 25\% of the stars in the sample with low-mass brown dwarfs
 companions are red giants. 
 To check for possible biases we have repeated the comparison of the abundance properties
 ([Fe/H], [X$_{\alpha}$/Fe], [X$_{\rm Fe}$/Fe], [X$_{\rm vol}$/Fe])
 of the SWBD with companions with masses above and below 42.5 M$_{\rm Jup}$,
 excluding from the analysis all stars classified as giants.
 The results are shown in Table~\ref{abundance_statistics3} where the new
 analysis is compared with the previous one. 
 It can be seen that the results do not change in a significant way.
 For example, for [Fe/H] the $p$-value changes from 0.08 to 0.05,
 while when considering [X$_{\rm Fe}$/Fe] it moves from less than 0.01 to 0.04.
 Although the threshold of 0.02 on the $p$-value is usually assumed to consider
 statistical significance when interpreting the results from the K-S tests,
 we note that a $p$-value of 0.04 is still very low. 
 We conclude that the presence of giant stars in the SWBDs with
 companions below 42.5 M$_{\rm Jup}$ does not introduce any significant
 bias in the comparisons performed in this work.

  However, the results might change if in addition to the giant stars
  we also exclude the subgiant stars
  (from both SWBDs subsamples). 
    In this case, the $p$-value for [Fe/H] increases from 0.08 up to 0.24,
  while for [X$_{\rm Fe}$/Fe] it rises from less than 0.01 up to a value of 0.64.
  This is in contrast to what we found when excluding only the giant
  stars from the analysis and may, at least partially, be due to the
  significant reduction of the sample size.   
  Note that by excluding both giant and subgiant stars from the 
  analysis we are reducing the sample to approximately half the original
  size.

 Finally, we analyse the results  when the stars with additional
 companions in the planetary mass are excluded
 (all of them in the sample of stars with low-mass brown dwarfs),
 see Table~\ref{abundance_statistics3}.
 In this case the significance of a possible metallicity difference
 between SWBDs with companions above and below 42.5 M$_{\rm Jup}$
 diminishes  (the $p$-value changes from 0.08 to 0.31).
 The $p$-value for [X$_{\rm Fe}$/Fe] also rises a bit from less than 0.01 to 0.03.
 We again conclude that no significant bias is introduced by
 the five SWBDs with companions below 42.5 M$_{\rm Jup}$ that, in addition
 to a brown dwarf companion, also harbours a companion in the gas-giant
 planetary mass domain.

\begin{table*}
\centering
\caption{
 Comparison between the abundances of stars with BDs with minimum
 masses M$_{\rm C}$$\sin i$ $>$ 42.5 M$_{\rm Jup}$
 and stars with BDs companions with minimum masses in the range
 M$_{\rm C}$$\sin i$ $<$ 42.5 M$_{\rm Jup}$ when: 
 i) all stars are included, ii) giant stars are excluded, 
 iii) giant and subgiant stars are excluded, and
 iv) stars with additional planetary companions are excluded.
}
\label{abundance_statistics3}
\begin{tabular}{lcccccccc}
\hline\noalign{\smallskip}
        &  \multicolumn{2}{c}{All}   & \multicolumn{2}{c}{Without}      & \multicolumn{2}{c}{Without}      &  \multicolumn{2}{c}{Without stars with} \\ 
        &  \multicolumn{2}{c}{stars} & \multicolumn{2}{c}{giant stars}  & \multicolumn{2}{c}{subgiant stars} & \multicolumn{2}{c}{gas-giant planets} \\ 
\hline
      &   $D$ & $p$-value          & $D$ & $p$-value            &  $D$ & $p$-value & $D$ & $p$-value \\
\hline
$[Fe/H]$      & 0.34  & 0.08     & 0.39  & 0.05 & 0.38 & 0.24 &	 0.27 & 0.31 \\      
$X_{\alpha}$  & 0.23  & 0.45     & 0.15  & 0.95 & 0.26 & 0.70 &	 0.26 & 0.33 \\
$X_{\rm Fe}$  & 0.47  & $<$ 0.01 & 0.40  & 0.04 & 0.27 & 0.64 &  0.40 & 0.03 \\
$X_{\rm vol}$ & 0.19  & 0.73     & 0.18  & 0.83 & 0.35 & 0.35 &  0.24 & 0.46 \\ 
\noalign{\smallskip}\hline\noalign{\smallskip}
\end{tabular}
\tablefoot{n$_{\rm eff}$ $\sim$ 12.7 (all stars); $\sim$ 11.2 (without giant stars);
$\sim$ 6.5 (without subgiant/giant stars); $\sim$ 11.8 (without planet hosts).}
\end{table*}

\subsubsection{Stellar distance bias}
\label{distance_bias}

 As shown in Sec.~\ref{possible_biases} our sample contains several stars far from the solar neighbourhood
 including objects up to distances of 2174 pc.
 However, most of the studies of the solar neighbourhood are volume limited. In particular
 it should be noticed that at the distance increases astrometry becomes difficult and therefore only
 minimum masses are available. 

 In order to check whether our results are affected or not by having stars at relatively large
 distances we have repeated the statistical analysis performed before by considering
 only the stars with distances lower than 50 pc and the stars located within 75 pc.
 Approximately 60\% of our stars are within 50 pc, while this percentage increases up to
 $\sim$ 77\% for a distance of 75 pc.
 The results are shown in Table~\ref{abundance_distance}, 
 and can be compared with the first column of Table~\ref{abundance_statistics3}.
 We find that the values of the KS statistic ($D$) do not change in a significant
 way. Regarding the $p$-values, only the ones corresponding to
 $X_{\alpha}$ and $X_{\rm vol}$ seem to increase 
 when the SWBD sample is limited to stars within 75 pc. 
 The interpretation, however, does not change: differences
 in metallicity and iron-peak elements seem to be present 
 (note the very low $p$-values)  between SWBDs with companions with  minimum masses
 above and below 42.5  M$_{\rm Jup}$ irrespectively of whether all stars
 or a volume-limited sample is considered.

\begin{table}
\centering
\caption{
 Comparison between the abundances of stars with BDs with minimum
 masses M$_{\rm C}$$\sin i$ $>$ 42.5 M$_{\rm Jup}$
 and stars with BDs companions with minimum masses in the range
 M$_{\rm C}$$\sin i$ $<$ 42.5 M$_{\rm Jup}$ when: 
 i) only stars up to 50 pc are considered, and ii) only stars up to 
 75 pc are considered. 
}
\label{abundance_distance}
\begin{tabular}{lcccc}
\hline\noalign{\smallskip}
        &  \multicolumn{2}{c}{d $<$ 50 pc}   & \multicolumn{2}{c}{d $<$ 75 pc}      \\ 
\hline
        &   $D$              &   $p$-value   &        $D$ &         $p$-value       \\
\hline
$[Fe/H]$      & 0.48  & 0.04     & 0.43  & 0.03 \\
$X_{\alpha}$  & 0.27  & 0.55     & 0.17  & 0.90 \\
$X_{\rm Fe}$  & 0.38  & 0.16     & 0.39  & 0.06 \\
$X_{\rm vol}$ & 0.32  & 0.34     & 0.15  & 0.97 \\
\noalign{\smallskip}\hline\noalign{\smallskip}
\end{tabular}
\tablefoot{n$_{\rm eff}$ $\sim$ 7.7 (d $<$ 50 pc); $\sim$ 10.2 (d $<$ 75 pc).}
\end{table}
\subsubsection{Minimum and true masses}
\label{minimum_mass}

 Another source of bias that might be influencing this study
 comes from the fact that 
 from most of our SWBDs only the minimum mass of the mass brown candidate
 companion is known. This is an important effect as the distribution of minimum masses
 given by radial  velocity surveys of brown dwarfs might be less indicative of a true substellar
 mass than for objects in the planetary mass regime
 \citep[see e.g.][]{2013PASP..125..933S}.

 Given that we do not have information
 regarding the inclination angle of the BD stellar systems
 we have tried to account for this effect by considering a series of scenarios:
 i) a ``pessimistic'' scenario in which all our stars are seeing at very low
 inclinations (15 degrees);
 ii) a ``favourable'' case in which all our stars  are seeing at high inclinations (85 degrees); 
 iii) 
  a random distribution $P(i)$ for the orientation of the inclination
  expressed as $P(i) di = \sin(i) di$. The average value
  of $\sin i$ assuming a random inclination, $\langle \sin(i) \rangle=0.785$, is then used to estimate the mass
  of the brown dwarfs. Although more complex algorithms exist to 
  compute the probability distribution for $\sin(i)$, it has been shown
  that the use of the average value produces
  similar results for small number statistics \citep[see][and references therein]{2006ApJ...640.1051G};
  
 and iv) performing a series of 10$^{\rm 4}$ simulations with random inclinations for each star.
 In all cases we keep the ``true'' brown-dwarf masses when available.
 
 Table~\ref{table_minimum_mass} shows the results from the KS test for all these scenarios.
 The conclusion is that unless we are in the unlikely case that most of the stars are
 seen at very low inclinations angles (case i) the results do not change in a significant
 way (compare with first column in Table~\ref{abundance_statistics3}). In particular, we note that the results
 from the scenario iii) 
 are very similar
 to the results without assuming any inclination.
 The results from the simulations performed in case iv) are somehow inconclusive  
 given the large spread found for the $p$-values.

\begin{table*}
\centering
\caption{
 Comparison between the abundances of stars with BDs with 
 masses above and below 42.5 M$_{\rm Jup}$ assuming when 
 only  M$_{\rm C}$$\sin i$ measurements
 (i.e. without a determined true mass) are available:
 i)  an inclination angle of 15 degrees, ii)  an inclination
 angle of 85 degrees, iii) 
   the average value
 of $\sin i$ assuming a random inclination, $\langle \sin(i) \rangle =0.785$,
 iv)  random inclinations (the mean values of 10$^{\rm 4}$ simulations  
 are shown with their corresponding standard deviations). }
\label{table_minimum_mass}
\begin{tabular}{lcccccccc}
\hline\noalign{\smallskip}
& \multicolumn{2}{c}{$i$ = 15$^{o}$} & \multicolumn{2}{c}{$i$ = 85$^{o}$} & \multicolumn{2}{c}{ $\langle \sin(i) \rangle$} & \multicolumn{2}{c}{$i$ random} \\
\hline
              & $D$ & $p$-value  & $D$ & $p$-value   &  $D$ & $p$-value & $D$  & $p$-value \\
\hline
$[Fe/H]$      & 0.31  & 0.52     & 0.34  & 0.08     & 0.34 & 0.11      & 0.32 $\pm$ 0.09 & 0.32 $\pm$ 0.27  \\
$X_{\alpha}$  & 0.29  & 0.61     & 0.23  & 0.45     & 0.18 & 0.80      & 0.23 $\pm$ 0.07 & 0.62 $\pm$ 0.29  \\
$X_{\rm Fe}$  & 0.37  & 0.31     & 0.47  & $<$ 0.01 & 0.44 & $<$ 0.01  & 0.37 $\pm$ 0.10 & 0.22 $\pm$ 0.24  \\
$X_{\rm vol}$ & 0.36  & 0.35     & 0.19  & 0.73     & 0.20 & 0.69      & 0.24 $\pm$ 0.07 & 0.58 $\pm$ 0.27  \\
\noalign{\smallskip}\hline\noalign{\smallskip}
\end{tabular}
\tablefoot{n$_{\rm eff}$ $\sim$ 5.9 ($i$ = 15$^{o}$); $\sim$ 12.7 ($i$ = 85$^{o}$); $\sim$ 11.8 ( $\langle \sin(i) \rangle$);
$\sim$ 9.5 $\pm$ 0.9 ($i$ random)}
\end{table*}

\subsection{Abundances and brown dwarfs properties}
\label{abun_bd_properties}

\begin{figure*}[!htb]
\centering
\begin{minipage}{0.32\linewidth}
\includegraphics[angle=270,scale=0.39]{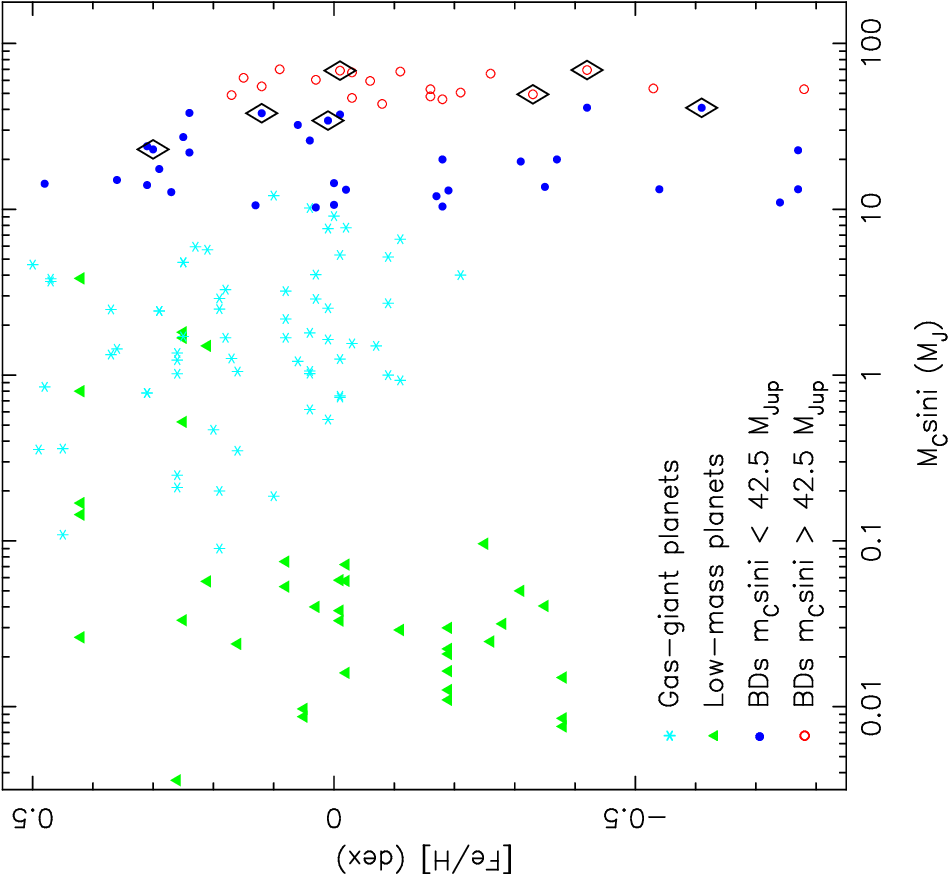}
\end{minipage}
\begin{minipage}{0.32\linewidth}
\includegraphics[angle=270,scale=0.39]{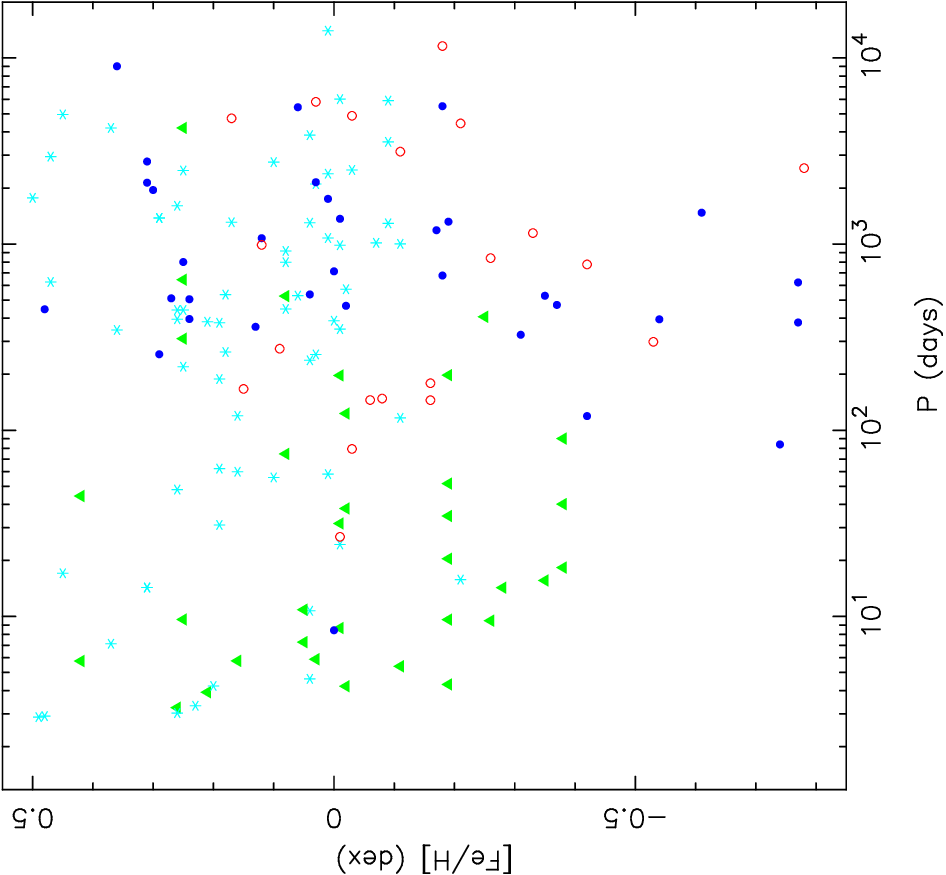}
\end{minipage}
\begin{minipage}{0.32\linewidth}
\includegraphics[angle=270,scale=0.39]{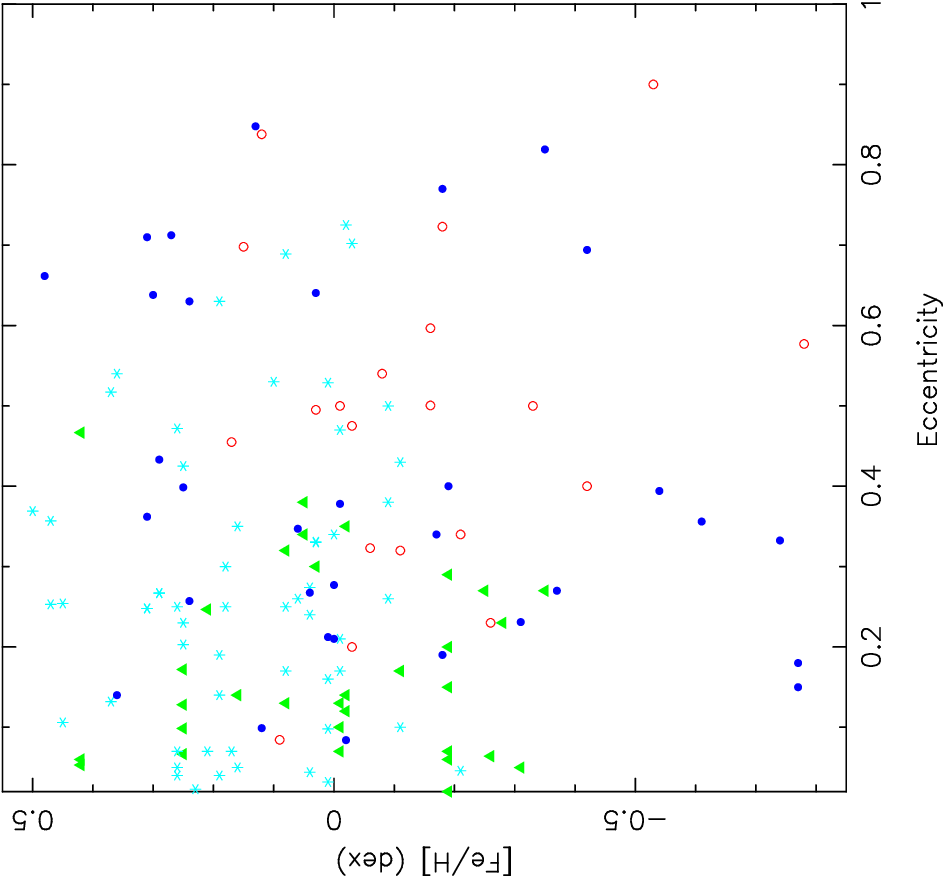}
\end{minipage}
\caption{
 Stellar metallicity as a function of the brown dwarf or planetary companion
 properties. 
 Colours and symbols are like in previous figures.
 Diamonds indicate brown dwarfs with ``true'' masses determinations.
 Left: minimum mass; Middle: period; Right: eccentricity.
}
\label{metallicity_bd_properties}
\end{figure*}

 A study of the possible relationships between stellar metallicity
 and the properties of the BD companions
 was also performed.
 Figure~\ref{metallicity_bd_properties} shows the stellar metallicity
 as a function of the BD minimum mass, period, and eccentricity.
 The figure does not reveal any clear correlation between the metallicity and the BDs
 properties. 

 The figure clearly shows the brown dwarf desert,
 as nearly 81.5\% of the BDs have periods larger than
 200 days. This is in sharp contrast with the presence
 of a significant number of gas-giant and low-mass planets
 at short periods. Among the stars with periods shorter than
 200 days, we note that only three BDs are in the mass range
 M$_{\rm C}$$\sin i$ $<$ 42.5 M$_{\rm Jup}$.
 Regarding the eccentricities, we note that BDs in
 the mass range  M$_{\rm C}$$\sin i$ $<$ 42.5 M$_{\rm Jup}$ tend
 to show low values, with  $\sim$ 70\% of the BDs in this
 mass range having eccentricities lower than 0.5.

\begin{figure*}[!htb]
\centering
\begin{minipage}{0.49\linewidth}
\includegraphics[angle=270,scale=0.45]{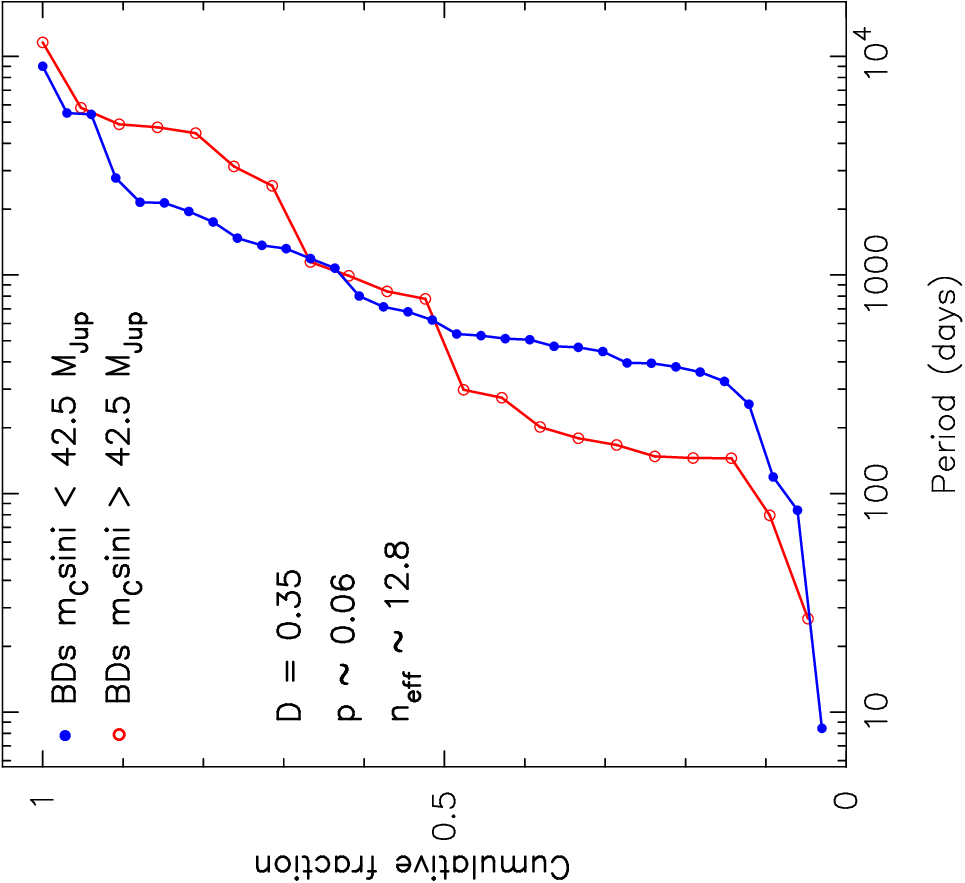}
\end{minipage}
\begin{minipage}{0.49\linewidth}
\includegraphics[angle=270,scale=0.45]{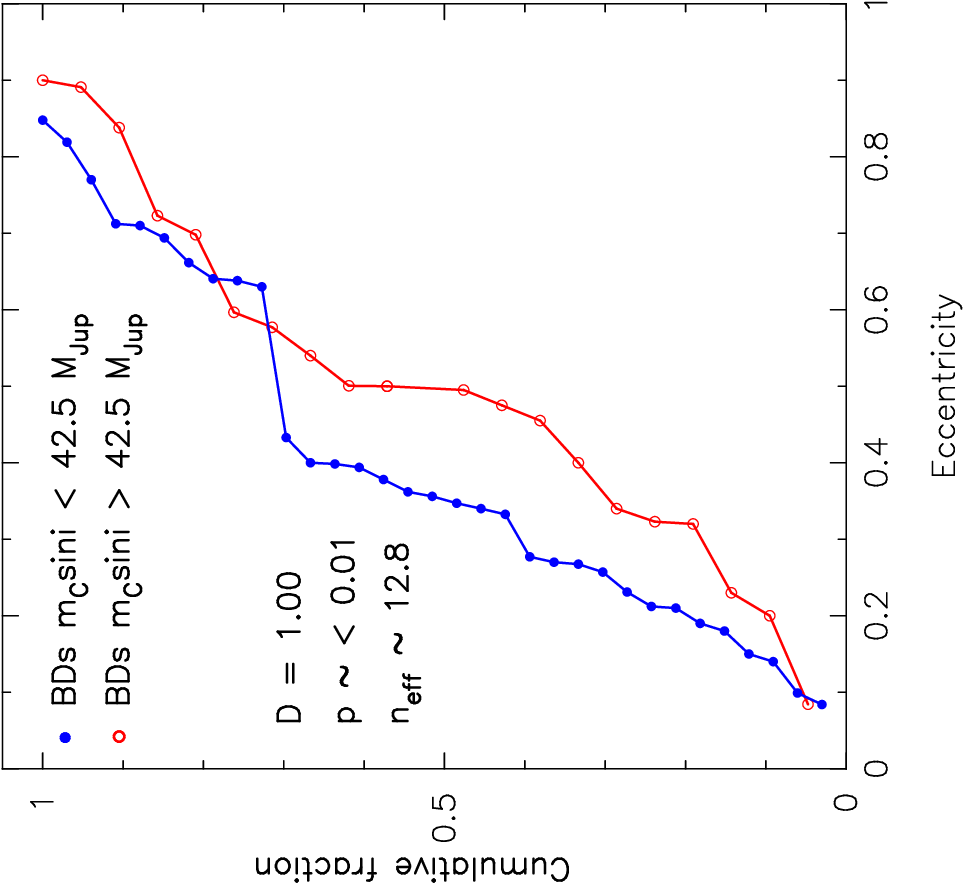}
\end{minipage}
\caption{
 Cumulative distribution function of periods (left) and eccentricities (right)
 for SWBDs with M$_{\rm C}$$\sin i$ $<$ 42.5 M$_{\rm Jup}$ and M$_{\rm C}$$\sin i$ $>$ 42.5 M$_{\rm Jup}$. }
\label{yd_old_comparison}
\end{figure*}

 Figure~\ref{yd_old_comparison} shows the cumulative distribution
 function of periods (left) and eccentricities (right) for
 SWBDs according to the mass of the companion.
 The analysis of the periods reveals that for periods shorter
 than $\sim$ 1000 days, the sample of SWBDs with M$_{\rm C}$$\sin i$ $>$ 42.5 M$_{\rm Jup}$
 shows shorter values than SWBDs with less massive companions.
 A K-S test gives a probability of both samples showing the
 same period distribution of $\sim$ 6\%.
 The sample of SWBDs with companions more massive than 42.5 M$_{\rm Jup}$ clearly
 shows higher eccentricities than the SWBDs with companions below 42.5 M$_{\rm Jup}$, 
 at least up to a value
 of $\sim$ 0.6.
 The K-S test on the eccentricity values suggets that both samples are statistically
 different ($p$-value $\sim$ 10$^{\rm -16}$).
 These results  are consistent 
  with the findings of
 \cite{2014MNRAS.439.2781M}.

\section{Discussion}\label{discusion}

 The existence of the brown dwarf desert has lead to numerous
 theories about whether brown dwarfs form like low-mass
 stars, like giant-planets or by entirely different mechanisms
 \citep[see e.g.][for a recent review]{2014prpl.conf..619C}. 
 The first observational  results of this work suggests that BDs
 should form in a different way from gas-giant planets (if metallicity as often assumed
 traces the formation mechanism), as
 it is clear from  Figure~\ref{distribuciones_acumuladas} that
 SWBDs do not follow the well-known gas-giant planet
 metallicity correlation. This can also be seen
 in the left panel of Figure~\ref{metallicity_bd_properties}.

 In a recent work, \cite{2014MNRAS.439.2781M}
 show that massive and low-mass brown dwarfs 
 have significantly different eccentricity distributions.
 This difference is also seen in our sample. 
 In particular, the authors note that BD with masses
 above 42.5 M$_{\rm Jup}$ have an eccentricity distribution consistent with that
 of binaries. 
 This result alone could be interesting in revealing clues regarding the formation mechanism of brown dwarfs.
 However, based alone on the eccentricity distribution
 we cannot directly infer that BDs and low mass stars are formed from the same process,
 i.e fragmentation of a molecular cloud. 
 Different formation mechanisms can lead to similar eccentricity distributions when subject
 to particular dynamical histories. What adds support to the hypothesis of a similar formation process
 is the fact that in our analysis we do not find any hint
 of metal enrichment in the stars with brown dwarf companions with masses above 
 42.5 M$_{\rm Jup}$. Moreover, in all the analysis performed in this work
 SWBDs with masses above 42.5 M$_{\rm Jup}$
 follow similar distributions to those of SWOPs or SWLMPs
 (see Figures~\ref{distribuciones_acumuladas}, and ~\ref{distrib_acu_alpha_iron_vol}),
 suggesting 
 a non-metallicity/abundance dependent formation.

 It has been shown \citep{2009MNRAS.392..413S} that BDs can be formed
 via gravitational instability in the outer parts ($>$ 100 au) of massive
 circumstellar discs (with stellar/disc mass ratios of the order of unity).
 The eccentricies were found to be very high as a result of this formation process
 but noted that just might be an artifact of the simulations that do not include
 tidal interactions with the gas disc. BDs in the so-called ejection 
 scenario are formed by gravoturbulent fragmentation of collapsing pre-stellar
 cores that due to dynamical interactions end-up being ejected from the cloud, terminating
 the accretion process 
 \citep[see i.e.][]{2009MNRAS.392..590B,2009MNRAS.397..232B}.
 In this later scenario eccentries are not expected to populate the high end of the eccentricity distribution.

 Support for a different formation mechanism for low-mass
 and massive brown dwarfs came from the chemical analysis performed
 in this work.
 Our results show  a tendency of SWBDs with masses below 42.5 M$_{\rm Jup}$
 of having slightly larger metallicities and abundances
  (especially X$_{\rm Fe}$) when compared with SWBDs with masses above 42.5 M$_{\rm Jup}$
 (see Figures~\ref{distribuciones_acumuladas},
 ~\ref{distribuciones_acumuladas_abundancias}, and ~\ref{distrib_acu_alpha_iron_vol})
 although with ``low'' 
  statistical significance
  (Table~\ref{abundance_statistics}) 
 We should note, however, that the results for X$_{\rm Fe}$
  are statistically significant.
 These results can be compared with the recent work
 by \cite{2014A&A...566A..83M}, where the authors already
 noticed the possible higher $\alpha$-element and Fe-peak abundances
 in the stars hosting brown dwarfs with masses below 42.5M $_{\rm Jup}$ (in comparison with
 those hosting more massive brown dwarfs), however these authors do not directly test the significance of these possible trends.
 Furthermore, their sample of SWBDs is significantly smaller than the
 one analysed in this work.

 If low-mass brown dwarfs were formed by core-accretion rather by
 the gravitational instability mechanism,
 stars hosting low-mass brown dwarfs should 
 show the metal-rich
 signature seen in gas-giant planetary hosts
 \citep[e.g.][]{1997MNRAS.285..403G,2004A&A...415.1153S,2005ApJ...622.1102F}.
 It is clear from Figure~\ref{distribuciones_acumuladas}, that SWBDs
 with masses below 42.5 M$_{\rm Jup}$ show lower metallicities
 than SWGPs. A K-S test confirms that both samples are different
 ($D$ $\sim$ 0.33, $p$-value $\sim$ 0.03, n$_{\rm eff}$ $\sim$ 18.5).
 Only for  metallicities above $\sim$  +0.20 dex, the metallicity distribution
 of SWBDs with masses below 42.5 M$_{\rm Jup}$ approaches the
 distribution of the SWGP sample.
 As already discussed, this might be affected by the presence of 
 additional planetary companions in some SWBDs.
 Indeed, when the stars with additional planets are removed
 from the SWBD sample, the higher metallicities of SWGPs 
 becomes more significant  ($D$ $\sim$ 0.40, $p$-value $\sim$ 0.007, n$_{\rm eff}$ $\sim$ 16.7).

 Low-mass brown dwarfs might form in self-gravitating protostellar discs \citep{2003MNRAS.346L..36R},
 a fast mechanism that does not requiere the previous formation of a rocky core and therefore
 it is independent of the stellar metallicity \citep{2002ApJ...567L.149B,2006ApJ...643..501B}. The simulations by
 \citet{2003MNRAS.346L..36R} shows that the fragmentation of an unstable protostellar disc produce a large number of
 substellar objects, although most of them are ejected from the system. 
 The remaining objects are typically either a very massive planet or a low-mass
 brown dwarf, having large periods and eccentricities\footnote {This is not at odds with our
 results from Figure~\ref{yd_old_comparison}, right panel, as usually eccentricities of the order of 0.2 are considered
 as ``large'' in the literature.}. It is possible, that four of the systems discussed in Section~\ref{possible_biases}
 (namely HD 38529, HD 168443, HIP 5158, and HAT-P-13)
 with a planet in an inner orbit and a brown dwarf at a larger distance formed in this way,
 as well as the two brown dwarf system around the metal-poor star BD+20 2457 ([Fe/H]=-0.77 dex).
 The case of the system around HD 202206 (also mentioned in Section~\ref{possible_biases})
 might need further discussion as the brown dwarf has
 an inner orbit to the planet one. 

 \cite{2003MNRAS.339.1025R} also shows that as the disc masses increases various effects might act
 to make the disc more unstable. 
 A relationship between the disc mass and the stellar mass of the form
 M$_{\rm disc}$ $\propto$ M$_{\star}^{1.2}$ 
 were suggested \citep{2011A&A...526A..63A} to explain the observed correlation between
 mass-accretion rate scales and stellar mass in young low-mass objects
 \citep{2003ApJ...592..266M,2004A&A...424..603N,2011A&A...535A..99M,2012A&A...543A..59M}.
 The fact that more massive stars might have more massive and more
 unstable discs might explain the presence of a relatively large number of low-mass BDs
 around evolved (subgiant and red giant) stars as shown that those are indeed more  massive stars
 \citep[e.g.][]{2013A&A...554A..84M}.

 So our results on the chemical analysis of BDs suggest that at low metallicities
 the dominant mechanism of BD formation is compatible with gravitational instability
 in massive discs or gravoturbulent fragmentation of collapsing pre-stellar cores
 (i.e. physical mechanisms that are not depend on the metal content of the cloud).
 The fact that we observed differences in the metal content for low and high mass BDs at high metallicities
 could indicate different mechanisms operating at different efficiencies. Core accretion
 might favor the formation of low-mass BDs at high metallicities even at low disc masses
 while inhibiting the formation of massive BDs  as not enough mass reservoir is available in the disc.
 For low-mass BDs  orbiting high-metallicity host stars the core acretion
 model might become efficient and favor the formation of
 BDs even at lower disc masses and inhibit the formation of BDs
 with larger masses (not enough mass in the disc).
 It is important to note that different BD/planet formation mechanisms can operate
 together and do not have to be exclusive of each other.

\section{Conclusions}\label{conclusions}

 In this work, a detailed chemical analysis of a large sample of
 stars with brown dwarfs has been presented.
 The sample has been analysed  taking into account the presence
 of massive (M$_{\rm C}$$\sin i$ $>$ 42.5 M$_{\rm Jup}$) and low-mass brown dwarfs
 (M$_{\rm C}$$\sin i$ $<$ 42.5 M$_{\rm Jup}$) companions. Before comparing both subsamples,
 a detailed analysis of their stellar
 properties was performed to control any possible bias affecting 
 our results. The chemical abundances of the SWBDs have also been compared to
 those of stars with known planetary companions as well as with 
 a sample of stars without planets.

 Our results show that SWBDs do not follow the well-known gas-giant
 metallicity correlation seen in main-sequence stars with planets.
 A tendency of SWBDs with substellar companions in the mass range
 M$_{\rm C}$$\sin i$ $>$ 42.5 M$_{\rm Jup}$ of having slightly
 larger metallicities and abundances than those of
 SWBDs with substellar companions in the mass range M$_{\rm C}$$\sin i$ $<$ 42.5 M$_{\rm Jup}$
 seems to be present in the data. However its statistical 
 significancy is rather low.
 We also confirm possible differences between SWBDs with substellar
 companions with masses above and below 42.5 M$_{\rm Jup}$ in terms
 of periods and eccentricities.
 All this observational evidence suggests that 
 the efficiencies of the different formation mechanisms
 may differ for low-mass and high-mass brown dwarfs.

 Our results are well described in a scenario in which
 high-mass brown dwarfs are mainly formed like low-mass stars (by the fragmention
 of a molecular cloud). Our analysis shows that
 at high metallicities the core-accretion model might be the
 mechanism for the formation of low-mass BDs.
 On the other hand, it seems reasonable that the most suitable
 scenario 
 for the formation of low-metallicity, low-mass BDs 
 is gravitational instability in turbulent protostellar discs
  since this mechanism is known to be independent of the 
  stellar metallicity.

\begin{acknowledgements}

  This research was supported by the Italian Ministry of Education,
  University, and Research  through the
  \emph{PREMIALE WOW 2013} research project under grant
  \emph{Ricerca di pianeti intorno a stelle di piccola massa}.
  E. V. acknowledges support from the \emph{On the rocks} project
  funded by the Spanish Ministerio de Econom\'ia y Competitividad
  under grant \emph{AYA2014-55840-P}.
  Carlos Eiroa is acknowledged for valuable discussions.
  We sincerely appreciate the careful
  reading of the manuscript and the constructive comments of an anonymous
  referee.

\end{acknowledgements}

%
\bibliographystyle{aa}
\bibliography{emarrones.bib}

\begin{thebibliography}{95}
\expandafter\ifx\csname natexlab\endcsname\relax\def\natexlab#1{#1}\fi

\bibitem[{{Adibekyan} {et~al.}(2012){Adibekyan}, {Sousa}, {Santos}, {Delgado
  Mena}, {Gonz{\'a}lez Hern{\'a}ndez}, {Israelian}, {Mayor}, \&
  {Khachatryan}}]{2012A&A...545A..32A}
{Adibekyan}, V.~Z., {Sousa}, S.~G., {Santos}, N.~C., {et~al.} 2012, \aap, 545,
  A32

\bibitem[{{Alibert} {et~al.}(2004){Alibert}, {Mordasini}, \&
  {Benz}}]{2004A&A...417L..25A}
{Alibert}, Y., {Mordasini}, C., \& {Benz}, W. 2004, \aap, 417, L25

\bibitem[{{Alibert} {et~al.}(2011){Alibert}, {Mordasini}, \&
  {Benz}}]{2011A&A...526A..63A}
{Alibert}, Y., {Mordasini}, C., \& {Benz}, W. 2011, \aap, 526, A63

\bibitem[{{Allende Prieto} {et~al.}(2001){Allende Prieto}, {Lambert}, \&
  {Asplund}}]{2001ApJ...556L..63A}
{Allende Prieto}, C., {Lambert}, D.~L., \& {Asplund}, M. 2001, \apjl, 556, L63

\bibitem[{{Arenou} {et~al.}(1992){Arenou}, {Grenon}, \&
  {Gomez}}]{1992A&A...258..104A}
{Arenou}, F., {Grenon}, M., \& {Gomez}, A. 1992, \aap, 258, 104

\bibitem[{{Baranne} {et~al.}(1996){Baranne}, {Queloz}, {Mayor}, {Adrianzyk},
  {Knispel}, {Kohler}, {Lacroix}, {Meunier}, {Rimbaud}, \&
  {Vin}}]{1996A&AS..119..373B}
{Baranne}, A., {Queloz}, D., {Mayor}, M., {et~al.} 1996, \aaps, 119, 373

\bibitem[{{Bate}(2009{\natexlab{a}})}]{2009MNRAS.392..590B}
{Bate}, M.~R. 2009{\natexlab{a}}, \mnras, 392, 590

\bibitem[{{Bate}(2009{\natexlab{b}})}]{2009MNRAS.397..232B}
{Bate}, M.~R. 2009{\natexlab{b}}, \mnras, 397, 232

\bibitem[{{Beir{\~a}o} {et~al.}(2005){Beir{\~a}o}, {Santos}, {Israelian}, \&
  {Mayor}}]{2005A&A...438..251B}
{Beir{\~a}o}, P., {Santos}, N.~C., {Israelian}, G., \& {Mayor}, M. 2005, \aap,
  438, 251

\bibitem[{{Bensby} {et~al.}(2003){Bensby}, {Feltzing}, \&
  {Lundstr{\"o}m}}]{2003A&A...410..527B}
{Bensby}, T., {Feltzing}, S., \& {Lundstr{\"o}m}, I. 2003, \aap, 410, 527

\bibitem[{{Bensby} {et~al.}(2005){Bensby}, {Feltzing}, {Lundstr{\"o}m}, \&
  {Ilyin}}]{2005A&A...433..185B}
{Bensby}, T., {Feltzing}, S., {Lundstr{\"o}m}, I., \& {Ilyin}, I. 2005, \aap,
  433, 185

\bibitem[{{Bergemann} {et~al.}(2011){Bergemann}, {Lind}, {Collet}, \&
  {Asplund}}]{2011JPhCS.328a2002B}
{Bergemann}, M., {Lind}, K., {Collet}, R., \& {Asplund}, M. 2011, Journal of
  Physics Conference Series, 328, 012002

\bibitem[{{Boss}(1997)}]{1997Sci...276.1836B}
{Boss}, A.~P. 1997, Science, 276, 1836

\bibitem[{{Boss}(2002)}]{2002ApJ...567L.149B}
{Boss}, A.~P. 2002, \apjl, 567, L149

\bibitem[{{Boss}(2006)}]{2006ApJ...643..501B}
{Boss}, A.~P. 2006, \apj, 643, 501

\bibitem[{{Bouchy} \& {Sophie Team}(2006)}]{2006tafp.conf..319B}
{Bouchy}, F. \& {Sophie Team}. 2006, in Tenth Anniversary of 51 Peg-b: Status
  of and prospects for hot Jupiter studies, ed. L.~{Arnold}, F.~{Bouchy}, \&
  C.~{Moutou}, 319--325

\bibitem[{{Bressan} {et~al.}(2012){Bressan}, {Marigo}, {Girardi}, {Salasnich},
  {Dal Cero}, {Rubele}, \& {Nanni}}]{2012MNRAS.427..127B}
{Bressan}, A., {Marigo}, P., {Girardi}, L., {et~al.} 2012, \mnras, 427, 127

\bibitem[{{Burgasser}(2011)}]{2011ASPC..450..113B}
{Burgasser}, A.~J. 2011, in Astronomical Society of the Pacific Conference
  Series, Vol. 450, Molecules in the Atmospheres of Extrasolar Planets, ed.
  J.~P. {Beaulieu}, S.~{Dieters}, \& G.~{Tinetti}, 113

\bibitem[{{Burrows} {et~al.}(2001){Burrows}, {Hubbard}, {Lunine}, \&
  {Liebert}}]{2001RvMP...73..719B}
{Burrows}, A., {Hubbard}, W.~B., {Lunine}, J.~I., \& {Liebert}, J. 2001,
  Reviews of Modern Physics, 73, 719

\bibitem[{{Burrows} {et~al.}(1997){Burrows}, {Marley}, {Hubbard}, {Lunine},
  {Guillot}, {Saumon}, {Freedman}, {Sudarsky}, \&
  {Sharp}}]{1997ApJ...491..856B}
{Burrows}, A., {Marley}, M., {Hubbard}, W.~B., {et~al.} 1997, \apj, 491, 856

\bibitem[{{Campbell} {et~al.}(1988){Campbell}, {Walker}, \&
  {Yang}}]{1988ApJ...331..902C}
{Campbell}, B., {Walker}, G.~A.~H., \& {Yang}, S. 1988, \apj, 331, 902

\bibitem[{{Casagrande} {et~al.}(2010){Casagrande}, {Ram{\'{\i}}rez},
  {Mel{\'e}ndez}, {Bessell}, \& {Asplund}}]{2010A&A...512A..54C}
{Casagrande}, L., {Ram{\'{\i}}rez}, I., {Mel{\'e}ndez}, J., {Bessell}, M., \&
  {Asplund}, M. 2010, \aap, 512, A54

\bibitem[{{Cassan} {et~al.}(2012){Cassan}, {Kubas}, {Beaulieu}, {Dominik},
  {Horne}, {Greenhill}, {Wambsganss}, {Menzies}, {Williams}, {J{\o}rgensen},
  {Udalski}, {Bennett}, {Albrow}, {Batista}, {Brillant}, {Caldwell}, {Cole},
  {Coutures}, {Cook}, {Dieters}, {Prester}, {Donatowicz}, {Fouqu{\'e}}, {Hill},
  {Kains}, {Kane}, {Marquette}, {Martin}, {Pollard}, {Sahu}, {Vinter},
  {Warren}, {Watson}, {Zub}, {Sumi}, {Szyma{\'n}ski}, {Kubiak}, {Poleski},
  {Soszynski}, {Ulaczyk}, {Pietrzy{\'n}ski}, \&
  {Wyrzykowski}}]{2012Natur.481..167C}
{Cassan}, A., {Kubas}, D., {Beaulieu}, J.-P., {et~al.} 2012, \nat, 481, 167

\bibitem[{{Chabrier} \& {Baraffe}(2000)}]{2000ARA&A..38..337C}
{Chabrier}, G. \& {Baraffe}, I. 2000, \araa, 38, 337

\bibitem[{{Chabrier} {et~al.}(2014){Chabrier}, {Johansen}, {Janson}, \&
  {Rafikov}}]{2014prpl.conf..619C}
{Chabrier}, G., {Johansen}, A., {Janson}, M., \& {Rafikov}, R. 2014, Protostars
  and Planets VI, 619

\bibitem[{{Cosentino} {et~al.}(2012){Cosentino}, {Lovis}, {Pepe}, {Collier
  Cameron}, {Latham}, {Molinari}, {Udry}, {Bezawada}, {Black}, {Born},
  {Buchschacher}, {Charbonneau}, {Figueira}, {Fleury}, {Galli}, {Gallie},
  {Gao}, {Ghedina}, {Gonzalez}, {Gonzalez}, {Guerra}, {Henry}, {Horne},
  {Hughes}, {Kelly}, {Lodi}, {Lunney}, {Maire}, {Mayor}, {Micela}, {Ordway},
  {Peacock}, {Phillips}, {Piotto}, {Pollacco}, {Queloz}, {Rice}, {Riverol},
  {Riverol}, {San Juan}, {Sasselov}, {Segransan}, {Sozzetti}, {Sosnowska},
  {Stobie}, {Szentgyorgyi}, {Vick}, \& {Weber}}]{2012SPIE.8446E..1VC}
{Cosentino}, R., {Lovis}, C., {Pepe}, F., {et~al.} 2012, in Society of
  Photo-Optical Instrumentation Engineers (SPIE) Conference Series, Vol. 8446,
  Society of Photo-Optical Instrumentation Engineers (SPIE) Conference Series,
  1

\bibitem[{{da Silva} {et~al.}(2006){da Silva}, {Girardi}, {Pasquini},
  {Setiawan}, {von der L{\"u}he}, {de Medeiros}, {Hatzes}, {D{\"o}llinger}, \&
  {Weiss}}]{2006A&A...458..609D}
{da Silva}, L., {Girardi}, L., {Pasquini}, L., {et~al.} 2006, \aap, 458, 609

\bibitem[{{Dekker} {et~al.}(2000){Dekker}, {D'Odorico}, {Kaufer}, {Delabre}, \&
  {Kotzlowski}}]{2000SPIE.4008..534D}
{Dekker}, H., {D'Odorico}, S., {Kaufer}, A., {Delabre}, B., \& {Kotzlowski}, H.
  2000, in \procspie, Vol. 4008, Optical and IR Telescope Instrumentation and
  Detectors, ed. M.~{Iye} \& A.~F. {Moorwood}, 534--545

\bibitem[{{Delgado Mena} {et~al.}(2010){Delgado Mena}, {Israelian},
  {Gonz{\'a}lez Hern{\'a}ndez}, {Bond}, {Santos}, {Udry}, \&
  {Mayor}}]{2010ApJ...725.2349D}
{Delgado Mena}, E., {Israelian}, G., {Gonz{\'a}lez Hern{\'a}ndez}, J.~I.,
  {et~al.} 2010, \apj, 725, 2349

\bibitem[{{Fischer} \& {Valenti}(2005)}]{2005ApJ...622.1102F}
{Fischer}, D.~A. \& {Valenti}, J. 2005, \apj, 622, 1102

\bibitem[{{Frandsen} \& {Lindberg}(1999)}]{1999anot.conf...71F}
{Frandsen}, S. \& {Lindberg}, B. 1999, in Astrophysics with the NOT, ed.
  H.~{Karttunen} \& V.~{Piirola}, 71

\bibitem[{{Ghezzi} {et~al.}(2010){Ghezzi}, {Cunha}, {Schuler}, \&
  {Smith}}]{2010ApJ...725..721G}
{Ghezzi}, L., {Cunha}, K., {Schuler}, S.~C., \& {Smith}, V.~V. 2010, \apj, 725,
  721

\bibitem[{{Gonzalez}(1997)}]{1997MNRAS.285..403G}
{Gonzalez}, G. 1997, \mnras, 285, 403

\bibitem[{{Grether} \& {Lineweaver}(2006)}]{2006ApJ...640.1051G}
{Grether}, D. \& {Lineweaver}, C.~H. 2006, \apj, 640, 1051

\bibitem[{{Hekker} \& {Mel{\'e}ndez}(2007)}]{2007A&A...475.1003H}
{Hekker}, S. \& {Mel{\'e}ndez}, J. 2007, \aap, 475, 1003

\bibitem[{{H{\o}g} {et~al.}(2000){H{\o}g}, {Fabricius}, {Makarov}, {Urban},
  {Corbin}, {Wycoff}, {Bastian}, {Schwekendiek}, \&
  {Wicenec}}]{2000A&A...355L..27H}
{H{\o}g}, E., {Fabricius}, C., {Makarov}, V.~V., {et~al.} 2000, \aap, 355, L27

\bibitem[{{Howard} {et~al.}(2013){Howard}, {Sanchis-Ojeda}, {Marcy}, {Johnson},
  {Winn}, {Isaacson}, {Fischer}, {Fulton}, {Sinukoff}, \&
  {Fortney}}]{2013Natur.503..381H}
{Howard}, A.~W., {Sanchis-Ojeda}, R., {Marcy}, G.~W., {et~al.} 2013, \nat, 503,
  381

\bibitem[{{Jofr{\'e}} {et~al.}(2015){Jofr{\'e}}, {Petrucci}, {Saffe}, {Saker},
  {de la Villarmois}, {Chavero}, {G{\'o}mez}, \& {Mauas}}]{2015A&A...574A..50J}
{Jofr{\'e}}, E., {Petrucci}, R., {Saffe}, C., {et~al.} 2015, \aap, 574, A50

\bibitem[{{Kaufer} {et~al.}(1999){Kaufer}, {Stahl}, {Tubbesing},
  {N{\o}rregaard}, {Avila}, {Francois}, {Pasquini}, \&
  {Pizzella}}]{1999Msngr..95....8K}
{Kaufer}, A., {Stahl}, O., {Tubbesing}, S., {et~al.} 1999, The Messenger, 95, 8

\bibitem[{{Kharchenko} {et~al.}(2007){Kharchenko}, {Scholz}, {Piskunov},
  {R{\"o}ser}, \& {Schilbach}}]{2007AN....328..889K}
{Kharchenko}, N.~V., {Scholz}, R.-D., {Piskunov}, A.~E., {R{\"o}ser}, S., \&
  {Schilbach}, E. 2007, Astronomische Nachrichten, 328, 889

\bibitem[{{Kurucz}(1993)}]{1993KurCD..13.....K}
{Kurucz}, R. 1993, ATLAS9 Stellar Atmosphere Programs and 2 km/s grid.~Kurucz
  CD-ROM No.~13.~ Cambridge, Mass.: Smithsonian Astrophysical Observatory,
  1993., 13

\bibitem[{{Luhman}(2012)}]{2012ARA&A..50...65L}
{Luhman}, K.~L. 2012, \araa, 50, 65

\bibitem[{{Luhman} {et~al.}(2007){Luhman}, {Joergens}, {Lada}, {Muzerolle},
  {Pascucci}, \& {White}}]{2007prpl.conf..443L}
{Luhman}, K.~L., {Joergens}, V., {Lada}, C., {et~al.} 2007, Protostars and
  Planets V, 443

\bibitem[{{Ma} \& {Ge}(2014)}]{2014MNRAS.439.2781M}
{Ma}, B. \& {Ge}, J. 2014, \mnras, 439, 2781

\bibitem[{{Maldonado} {et~al.}(2015){Maldonado}, {Eiroa}, {Villaver},
  {Montesinos}, \& {Mora}}]{2015A&A...579A..20M}
{Maldonado}, J., {Eiroa}, C., {Villaver}, E., {Montesinos}, B., \& {Mora}, A.
  2015, \aap, 579, A20

\bibitem[{{Maldonado} {et~al.}(2010){Maldonado}, {Mart{\'{\i}}nez-Arn{\'a}iz},
  {Eiroa}, {Montes}, \& {Montesinos}}]{2010A&A...521A..12M}
{Maldonado}, J., {Mart{\'{\i}}nez-Arn{\'a}iz}, R.~M., {Eiroa}, C., {Montes},
  D., \& {Montesinos}, B. 2010, \aap, 521, A12

\bibitem[{{Maldonado} \& {Villaver}(2016)}]{2016A&A...588A..98M}
{Maldonado}, J. \& {Villaver}, E. 2016, \aap, 588, A98

\bibitem[{{Maldonado} {et~al.}(2013){Maldonado}, {Villaver}, \&
  {Eiroa}}]{2013A&A...554A..84M}
{Maldonado}, J., {Villaver}, E., \& {Eiroa}, C. 2013, \aap, 554, A84

\bibitem[{{Marcy} \& {Butler}(2000)}]{2000PASP..112..137M}
{Marcy}, G.~W. \& {Butler}, R.~P. 2000, \pasp, 112, 137

\bibitem[{{Mashonkina} {et~al.}(2011){Mashonkina}, {Gehren}, {Shi}, {Korn}, \&
  {Grupp}}]{2011A&A...528A..87M}
{Mashonkina}, L., {Gehren}, T., {Shi}, J.-R., {Korn}, A.~J., \& {Grupp}, F.
  2011, \aap, 528, A87

\bibitem[{{Mata S{\'a}nchez} {et~al.}(2014){Mata S{\'a}nchez}, {Gonz{\'a}lez
  Hern{\'a}ndez}, {Israelian}, {Santos}, {Sahlmann}, \&
  {Udry}}]{2014A&A...566A..83M}
{Mata S{\'a}nchez}, D., {Gonz{\'a}lez Hern{\'a}ndez}, J.~I., {Israelian}, G.,
  {et~al.} 2014, \aap, 566, A83

\bibitem[{{Mayor} {et~al.}(2011){Mayor}, {Marmier}, {Lovis}, {Udry},
  {S{\'e}gransan}, {Pepe}, {Benz}, {Bertaux}, {Bouchy}, {Dumusque}, {Lo Curto},
  {Mordasini}, {Queloz}, \& {Santos}}]{2011arXiv1109.2497M}
{Mayor}, M., {Marmier}, M., {Lovis}, C., {et~al.} 2011, ArXiv e-prints

\bibitem[{{Mayor} {et~al.}(2003){Mayor}, {Pepe}, {Queloz}, {Bouchy},
  {Rupprecht}, {Lo Curto}, {Avila}, {Benz}, {Bertaux}, {Bonfils}, {Dall},
  {Dekker}, {Delabre}, {Eckert}, {Fleury}, {Gilliotte}, {Gojak}, {Guzman},
  {Kohler}, {Lizon}, {Longinotti}, {Lovis}, {Megevand}, {Pasquini}, {Reyes},
  {Sivan}, {Sosnowska}, {Soto}, {Udry}, {van Kesteren}, {Weber}, \&
  {Weilenmann}}]{2003Msngr.114...20M}
{Mayor}, M., {Pepe}, F., {Queloz}, D., {et~al.} 2003, The Messenger, 114, 20

\bibitem[{{Mendigut{\'{\i}}a} {et~al.}(2011){Mendigut{\'{\i}}a}, {Calvet},
  {Montesinos}, {Mora}, {Muzerolle}, {Eiroa}, {Oudmaijer}, \&
  {Mer{\'{\i}}n}}]{2011A&A...535A..99M}
{Mendigut{\'{\i}}a}, I., {Calvet}, N., {Montesinos}, B., {et~al.} 2011, \aap,
  535, A99

\bibitem[{{Mendigut{\'{\i}}a} {et~al.}(2012){Mendigut{\'{\i}}a}, {Mora},
  {Montesinos}, {Eiroa}, {Meeus}, {Mer{\'{\i}}n}, \&
  {Oudmaijer}}]{2012A&A...543A..59M}
{Mendigut{\'{\i}}a}, I., {Mora}, A., {Montesinos}, B., {et~al.} 2012, \aap,
  543, A59

\bibitem[{{Montes} {et~al.}(2001){Montes}, {L{\'o}pez-Santiago}, {G{\'a}lvez},
  {Fern{\'a}ndez-Figueroa}, {De Castro}, \& {Cornide}}]{2001MNRAS.328...45M}
{Montes}, D., {L{\'o}pez-Santiago}, J., {G{\'a}lvez}, M.~C., {et~al.} 2001,
  \mnras, 328, 45

\bibitem[{{Mordasini} {et~al.}(2012){Mordasini}, {Alibert}, {Benz}, {Klahr}, \&
  {Henning}}]{2012A&A...541A..97M}
{Mordasini}, C., {Alibert}, Y., {Benz}, W., {Klahr}, H., \& {Henning}, T. 2012,
  \aap, 541, A97

\bibitem[{{Mortier} {et~al.}(2013){Mortier}, {Santos}, {Sousa}, {Adibekyan},
  {Delgado Mena}, {Tsantaki}, {Israelian}, \& {Mayor}}]{2013A&A...557A..70M}
{Mortier}, A., {Santos}, N.~C., {Sousa}, S.~G., {et~al.} 2013, \aap, 557, A70

\bibitem[{{Murdoch} {et~al.}(1993){Murdoch}, {Hearnshaw}, \&
  {Clark}}]{1993ApJ...413..349M}
{Murdoch}, K.~A., {Hearnshaw}, J.~B., \& {Clark}, M. 1993, \apj, 413, 349

\bibitem[{{Muzerolle} {et~al.}(2003){Muzerolle}, {Hillenbrand}, {Calvet},
  {Brice{\~n}o}, \& {Hartmann}}]{2003ApJ...592..266M}
{Muzerolle}, J., {Hillenbrand}, L., {Calvet}, N., {Brice{\~n}o}, C., \&
  {Hartmann}, L. 2003, \apj, 592, 266

\bibitem[{{Natta} {et~al.}(2004){Natta}, {Testi}, {Muzerolle}, {Randich},
  {Comer{\'o}n}, \& {Persi}}]{2004A&A...424..603N}
{Natta}, A., {Testi}, L., {Muzerolle}, J., {et~al.} 2004, \aap, 424, 603

\bibitem[{{Neves} {et~al.}(2009){Neves}, {Santos}, {Sousa}, {Correia}, \&
  {Israelian}}]{2009A&A...497..563N}
{Neves}, V., {Santos}, N.~C., {Sousa}, S.~G., {Correia}, A.~C.~M., \&
  {Israelian}, G. 2009, \aap, 497, 563

\bibitem[{{Niedzielski} {et~al.}(2009){Niedzielski}, {Nowak}, {Adam{\'o}w}, \&
  {Wolszczan}}]{2009ApJ...707..768N}
{Niedzielski}, A., {Nowak}, G., {Adam{\'o}w}, M., \& {Wolszczan}, A. 2009,
  \apj, 707, 768

\bibitem[{{Nordstr{\"o}m} {et~al.}(2004){Nordstr{\"o}m}, {Mayor}, {Andersen},
  {Holmberg}, {Pont}, {J{\o}rgensen}, {Olsen}, {Udry}, \&
  {Mowlavi}}]{2004A&A...418..989N}
{Nordstr{\"o}m}, B., {Mayor}, M., {Andersen}, J., {et~al.} 2004, \aap, 418, 989

\bibitem[{{Pasquini} {et~al.}(2007){Pasquini}, {D{\"o}llinger}, {Weiss},
  {Girardi}, {Chavero}, {Hatzes}, {da Silva}, \&
  {Setiawan}}]{2007A&A...473..979P}
{Pasquini}, L., {D{\"o}llinger}, M.~P., {Weiss}, A., {et~al.} 2007, \aap, 473,
  979

\bibitem[{{Peacock}(1983)}]{1983MNRAS.202..615P}
{Peacock}, J.~A. 1983, \mnras, 202, 615

\bibitem[{{Perryman} \& {ESA}(1997)}]{1997ESASP1200.....P}
{Perryman}, M.~A.~C. \& {ESA}, eds. 1997, ESA Special Publication, Vol. 1200,
  {The HIPPARCOS and TYCHO catalogues. Astrometric and photometric star
  catalogues derived from the ESA HIPPARCOS Space Astrometry Mission}

\bibitem[{{Pollack} {et~al.}(1996){Pollack}, {Hubickyj}, {Bodenheimer},
  {Lissauer}, {Podolak}, \& {Greenzweig}}]{1996Icar..124...62P}
{Pollack}, J.~B., {Hubickyj}, O., {Bodenheimer}, P., {et~al.} 1996, \icarus,
  124, 62

\bibitem[{{Raskin} {et~al.}(2011){Raskin}, {van Winckel}, {Hensberge},
  {Jorissen}, {Lehmann}, {Waelkens}, {Avila}, {de Cuyper}, {Degroote},
  {Dubosson}, {Dumortier}, {Fr{\'e}mat}, {Laux}, {Michaud}, {Morren}, {Perez
  Padilla}, {Pessemier}, {Prins}, {Smolders}, {van Eck}, \&
  {Winkler}}]{2011A&A...526A..69R}
{Raskin}, G., {van Winckel}, H., {Hensberge}, H., {et~al.} 2011, \aap, 526, A69

\bibitem[{{Reffert} {et~al.}(2015){Reffert}, {Bergmann}, {Quirrenbach},
  {Trifonov}, \& {K{\"u}nstler}}]{2015A&A...574A.116R}
{Reffert}, S., {Bergmann}, C., {Quirrenbach}, A., {Trifonov}, T., \&
  {K{\"u}nstler}, A. 2015, \aap, 574, A116

\bibitem[{{Rice} \& {Armitage}(2003)}]{2003ApJ...598L..55R}
{Rice}, W.~K.~M. \& {Armitage}, P.~J. 2003, \apjl, 598, L55

\bibitem[{{Rice} {et~al.}(2003{\natexlab{a}}){Rice}, {Armitage}, {Bate}, \&
  {Bonnell}}]{2003MNRAS.339.1025R}
{Rice}, W.~K.~M., {Armitage}, P.~J., {Bate}, M.~R., \& {Bonnell}, I.~A.
  2003{\natexlab{a}}, \mnras, 339, 1025

\bibitem[{{Rice} {et~al.}(2003{\natexlab{b}}){Rice}, {Armitage}, {Bonnell},
  {Bate}, {Jeffers}, \& {Vine}}]{2003MNRAS.346L..36R}
{Rice}, W.~K.~M., {Armitage}, P.~J., {Bonnell}, I.~A., {et~al.}
  2003{\natexlab{b}}, \mnras, 346, L36

\bibitem[{{Sadakane} {et~al.}(2005){Sadakane}, {Ohnishi}, {Ohkubo}, \&
  {Takeda}}]{2005PASJ...57..127S}
{Sadakane}, K., {Ohnishi}, T., {Ohkubo}, M., \& {Takeda}, Y. 2005, \pasj, 57,
  127

\bibitem[{{Sahlmann} {et~al.}(2011){Sahlmann}, {S{\'e}gransan}, {Queloz},
  {Udry}, {Santos}, {Marmier}, {Mayor}, {Naef}, {Pepe}, \&
  {Zucker}}]{2011A&A...525A..95S}
{Sahlmann}, J., {S{\'e}gransan}, D., {Queloz}, D., {et~al.} 2011, \aap, 525,
  A95

\bibitem[{{Santos} {et~al.}(2004){Santos}, {Israelian}, \&
  {Mayor}}]{2004A&A...415.1153S}
{Santos}, N.~C., {Israelian}, G., \& {Mayor}, M. 2004, \aap, 415, 1153

\bibitem[{{Santos} {et~al.}(2013){Santos}, {Sousa}, {Mortier}, {Neves},
  {Adibekyan}, {Tsantaki}, {Delgado Mena}, {Bonfils}, {Israelian}, {Mayor}, \&
  {Udry}}]{2013A&A...556A.150S}
{Santos}, N.~C., {Sousa}, S.~G., {Mortier}, A., {et~al.} 2013, \aap, 556, A150

\bibitem[{{Schuler} {et~al.}(2005){Schuler}, {Kim}, {Tinker}, {King}, {Hatzes},
  \& {Guenther}}]{2005ApJ...632L.131S}
{Schuler}, S.~C., {Kim}, J.~H., {Tinker}, Jr., M.~C., {et~al.} 2005, \apjl,
  632, L131

\bibitem[{{Sneden}(1973)}]{1973PhDT.......180S}
{Sneden}, C.~A. 1973, PhD thesis, THE UNIVERSITY OF TEXAS AT AUSTIN.

\bibitem[{{Sousa} {et~al.}(2015){Sousa}, {Santos}, {Adibekyan}, {Delgado-Mena},
  \& {Israelian}}]{2015A&A...577A..67S}
{Sousa}, S.~G., {Santos}, N.~C., {Adibekyan}, V., {Delgado-Mena}, E., \&
  {Israelian}, G. 2015, \aap, 577, A67

\bibitem[{{Sousa} {et~al.}(2008){Sousa}, {Santos}, {Mayor}, {Udry},
  {Casagrande}, {Israelian}, {Pepe}, {Queloz}, \&
  {Monteiro}}]{2008A&A...487..373S}
{Sousa}, S.~G., {Santos}, N.~C., {Mayor}, M., {et~al.} 2008, \aap, 487, 373

\bibitem[{{Spiegel} {et~al.}(2011){Spiegel}, {Burrows}, \&
  {Milsom}}]{2011ApJ...727...57S}
{Spiegel}, D.~S., {Burrows}, A., \& {Milsom}, J.~A. 2011, \apj, 727, 57

\bibitem[{{Stamatellos} \& {Whitworth}(2009)}]{2009MNRAS.392..413S}
{Stamatellos}, D. \& {Whitworth}, A.~P. 2009, \mnras, 392, 413

\bibitem[{{Stevens} \& {Gaudi}(2013)}]{2013PASP..125..933S}
{Stevens}, D.~J. \& {Gaudi}, B.~S. 2013, \pasp, 125, 933

\bibitem[{{Takeda}(2007)}]{2007PASJ...59..335T}
{Takeda}, Y. 2007, \pasj, 59, 335

\bibitem[{{Takeda} {et~al.}(2002{\natexlab{a}}){Takeda}, {Ohkubo}, \&
  {Sadakane}}]{2002PASJ...54..451T}
{Takeda}, Y., {Ohkubo}, M., \& {Sadakane}, K. 2002{\natexlab{a}}, \pasj, 54,
  451

\bibitem[{{Takeda} {et~al.}(2005){Takeda}, {Ohkubo}, {Sato}, {Kambe}, \&
  {Sadakane}}]{2005PASJ...57...27T}
{Takeda}, Y., {Ohkubo}, M., {Sato}, B., {Kambe}, E., \& {Sadakane}, K. 2005,
  \pasj, 57, 27

\bibitem[{{Takeda} {et~al.}(2002{\natexlab{b}}){Takeda}, {Sato}, {Kambe},
  {Sadakane}, \& {Ohkubo}}]{2002PASJ...54.1041T}
{Takeda}, Y., {Sato}, B., {Kambe}, E., {Sadakane}, K., \& {Ohkubo}, M.
  2002{\natexlab{b}}, \pasj, 54, 1041

\bibitem[{{Takeda} {et~al.}(2008){Takeda}, {Sato}, \&
  {Murata}}]{2008PASJ...60..781T}
{Takeda}, Y., {Sato}, B., \& {Murata}, D. 2008, \pasj, 60, 781

\bibitem[{{Valenti} \& {Fischer}(2005)}]{2005ApJS..159..141V}
{Valenti}, J.~A. \& {Fischer}, D.~A. 2005, \apjs, 159, 141

\bibitem[{van Leeuwen(2007)}]{Leeuwen}
van Leeuwen, F.~v. 2007, Hipparcos, the New Reduction of the Raw Data (XXXII,
  449 p., Hardcover, ISBN: 978-1-4020-6341-1: Astrophysics and Space Science
  Library , Vol. 350)

\bibitem[{{Wenger} {et~al.}(2000){Wenger}, {Ochsenbein}, {Egret}, {Dubois},
  {Bonnarel}, {Borde}, {Genova}, {Jasniewicz}, {Lalo{\"e}}, {Lesteven}, \&
  {Monier}}]{2000A&AS..143....9W}
{Wenger}, M., {Ochsenbein}, F., {Egret}, D., {et~al.} 2000, \aaps, 143, 9

\bibitem[{{Whitworth} {et~al.}(2007){Whitworth}, {Bate}, {Nordlund},
  {Reipurth}, \& {Zinnecker}}]{2007prpl.conf..459W}
{Whitworth}, A., {Bate}, M.~R., {Nordlund}, {\AA}., {Reipurth}, B., \&
  {Zinnecker}, H. 2007, Protostars and Planets V, 459

\bibitem[{{Wilson} {et~al.}(2016){Wilson}, {H{\'e}brard}, {Santos}, {Sahlmann},
  {Montagnier}, {Astudillo-Defru}, {Boisse}, {Bouchy}, {Rey}, {Arnold},
  {Bonfils}, {Bourrier}, {Courcol}, {Deleuil}, {Delfosse}, {D{\'{\i}}az},
  {Ehrenreich}, {Forveille}, {Moutou}, {Pepe}, {Santerne}, {S{\'e}gransan}, \&
  {Udry}}]{2016A&A...588A.144W}
{Wilson}, P.~A., {H{\'e}brard}, G., {Santos}, N.~C., {et~al.} 2016, \aap, 588,
  A144

\bibitem[{{Zieli{\'n}ski} {et~al.}(2012){Zieli{\'n}ski}, {Niedzielski},
  {Wolszczan}, {Adam{\'o}w}, \& {Nowak}}]{2012A&A...547A..91Z}
{Zieli{\'n}ski}, P., {Niedzielski}, A., {Wolszczan}, A., {Adam{\'o}w}, M., \&
  {Nowak}, G. 2012, \aap, 547, A91

\end{thebibliography}


\end{document}